\documentclass[12pt]{article}
\usepackage{booktabs}
\usepackage{epsfig}
\usepackage{mycaption}
\usepackage{fancybox}
\usepackage{ourbox}
\newcounter{longversion}
\ifnum \arabic{longversion} > 0 
\usepackage{lsalikedjcm}
\else 
\newcommand{\citeasnoun}[1]{\cite{#1}\typeout{cite as noun #1}}
\newcommand{\quotecite}[1]{\cite{#1}'s\typeout{quotecite #1}}
\fi
\input{psfig.tex}
\newsavebox{\leftDbox}
\savebox{\leftDbox}{%
\setlength{\unitlength}{0.01cm}
\begin{picture}(36,36)(14,-18)
\put(50,0){\makebox(0,0){\makebox[0in][r]{$\langle$}{\sc{d}}}}
\put(65,18){\line(-1,0){27}}
\put(65,-18){\line(-1,0){27}}
\put(65,18){\line(0,-1){36}}
\end{picture}
}

\newcommand{\eqref}[1]{equation~(\ref{#1})}

\newcommand{\Eqref}[1]{Equation~(\ref{#1})}

\newcommand{\boxref}{box~\ref}

\newcommand{\figref}{figure~\ref}
\newcommand{\Figref}{Figure~\ref}

\newcommand{\tabref}{table~\ref}
\newcommand{\tablenoun}{table}

\newcounter{frompage}
\newcommand{\dvips}{\ifnum \arabic{frompage} < \arabic{page} 
\typeout{\# dvips Chapter \arabic{chapter} }
\typeout{ dvips -p \arabic{frompage} -l \arabic{page} -o ps/\arabic{frompage}.\arabic{page}.ps }
\setcounter{frompage}{\arabic{page}}
\addtocounter{frompage}{1}
\else
\typeout{ Already printed to here}
\fi
}
\newcommand{\dvipsb}[1]{\ifnum \arabic{frompage} < \arabic{page} 
\typeout{\# dvips #1}
\typeout{ dvips -p \arabic{frompage} -l \arabic{page} -o ps/\arabic{frompage}.\arabic{page}.ps }
\setcounter{frompage}{\arabic{page}}
\addtocounter{frompage}{1}
\else
\typeout{ Already printed to here}
\fi
}

\newcommand{\linefrac}[2]{{#1}/{#2}}

\newcommand{\noprint}[1]{}

\newlength{\xx}
\newlength{\xxx}
\newlength{\yy}
\newcommand{\bg}{{\bf g}}

\font\tinysf=cmss8 at 6pt 
\def\T{\hspace*{-0.2mm}\mbox{\tinysf T}\hspace*{-0.2mm}}

\renewcommand{\S}{{\cal S}}

\renewcommand{\P}{{\cal P}}

\newcommand{\bff}{{\bf f}}

\newcommand{\eg}{{\it e.g.}}
\newcommand{\ie}{{\it i.e.}}
\newcommand{\cf}{{\it cf.}}

\newcommand{\bC}{{\bf C}}

\newcommand{\bH}{{\bf H}}

\newcommand{\bR}{{\bf R}}

\newcommand{\bz}{{\bf z}}

\newcommand{\bI}{{\bf I}}

\newcommand{\be}{{\bf e}}

\newcommand{\bx}{{\bf x}}
\newcommand{\by}{{\bf y}}

\newcommand{\dfrac}[2]{\mbox{{\raisebox{3pt}{\footnotesize{#1}}}%
\hspace{-0.4mm}{\raisebox{2pt}{\small{/}}}\hspace{-0.4mm}%
{{\footnotesize{#2}}}}}
\newcommand{\mdfrac}[2]{\mbox{{\raisebox{3pt}{\footnotesize{$#1$}}}%
\hspace{-0.4mm}{\raisebox{2pt}{\small{/}}}\hspace{-0.4mm}%
{{\footnotesize{$#2$}}}}}

\newcommand{\dhalf}{\dfrac{1}{2}}
\newcommand{\half}{\frac{1}{2}}

\newcommand{\ben}{\begin{enumerate}} 
\newcommand{\een}{\end{enumerate}}
\newcommand{\beq}{\begin{equation}}
\newcommand{\eeq}{\end{equation}}
\newcommand{\beqa}{\begin{eqnarray*}}
\newcommand{\eeqa}{\end{eqnarray*}}
\newcommand{\beqan}{\begin{eqnarray}}
\newcommand{\eeqan}{\end{eqnarray}}
\newcommand{\bt}{{\bf t}}
\newcommand{\btt}{\begin{tabular}}
\newcommand{\et}{\end{tabular}}

\newcommand{\bs}{{\bf s}}

\newcommand{\I}{{I}}

\newcounter{mycounter}
\newcommand\leftmarginpar[1]{%
        \stepcounter{mycounter}%
        \label{leftmargin:\arabic{mycounter}}
        \ifodd\mp@pageref{leftmargin:\arabic{mycounter}}
           \begingroup
           \reversemarginpar\marginpar{#1}%
           \endgroup
        \else
           \marginpar{#1}%
        \fi
}
\newcommand{\erf}{\Phi}
\newcommand{\ket}[1]{\ensuremath{\left | #1 \right \rangle}}

\newcommand{\eq}{\begin{equation}}
\newcommand{\en}{\end{equation}}

\newcommand{\operatorfont}{\tt}
\newcommand{\operatorfontb}{\ensuremath}
\newcommand{\X}{{\operatorfont X}}
\newcommand{\Y}{{\operatorfont Y}}
\newcommand{\Z}{{\operatorfont Z}}
\renewcommand{\I}{{\operatorfont I}}
\newcommand{\E}{{\operatorfontb E}}
\renewcommand{\O}{{\operatorfontb O}}
\renewcommand{\P}{{\operatorfontb P}}

\newcommand{\K}{K}
\renewcommand{\S}{{\operatorfontb S}}
\newcommand{\MM}{\ensuremath{{\bf A}}}
\renewcommand{\AA}{{\ensuremath{\bf H}}}
\renewcommand{\AA}{{{\bf H}}}
\newcommand{\AAA}{{{\bf A}_2}}
\newcommand{\BB}{{\ensuremath{{\bf G}}}}
\newcommand{\BBB}{{\ensuremath{{\bf A}_1}}}%
\newcommand{\tU}{\ensuremath{{\operatorfont U}}}%
\newcommand{\tM}{\ensuremath{{\operatorfont M}}}%
\newcommand{\tP}{\ensuremath{{\operatorfont P}}}%
\newcommand{\tQ}{\ensuremath{{\operatorfont Q}}}%
\newcommand{\PP}{\ensuremath{{\bf P}}}%
\newcommand{\QQ}{\ensuremath{{\bf Q}}}%
\newcommand{\bi}{{{\bf i}}}%

\newcommand{\bit}{\begin{itemize}}
\newcommand{\eit}{\end{itemize}} 

\renewcommand{\bs}{{\bf s}}

\newcommand{\noise}{e}
\renewcommand{\bff}{{\bf f}}
\renewcommand{\bH}{{\ensuremath{\bf H}}}

\renewcommand{\bx}{\ensuremath{\bf x}}
\renewcommand{\bz}{\ensuremath{\bf z}}

\newcommand{\secref}[1]{section~\ref{#1}}

\newcommand{\ebnobig}[1]{\begin{tabular}{r}%
\mbox{\hspace*{0.2in}\psfig{figure=figs/#1.ps,width=4.15in,angle=-90}}\\
$\fm$\\
\end{tabular}
}
\newcommand{\ebno}[1]{\mbox{\hspace*{0.2in}\psfig{figure=figs/#1.ps,width=2.15in,angle=-90}}}

\newcommand{\fm}{\ensuremath{f_{\rm m}}}
\newcommand{\ldpcc}{low-density parity-check code}

\newlength{\mylength}
\newcommand{\resetfboxsep}{\setlength{\fboxsep}{3pt}}
\newcommand{\fatfboxsep}{\setlength{\fboxsep}{10pt}}
\newenvironment{framedalgorithmw}[1]{
\fatfboxsep\par\noindent\begin{Sbox}%
\setlength{\mylength}{#1}%
\addtolength{\mylength}{-2\fboxsep}%
\addtolength{\mylength}{-2\fboxrule}%
\begin{minipage}{\mylength}}%
{\end{minipage}\end{Sbox}\fbox{\TheSbox}\resetfboxsep}

\begin{document}
\title{Sparse-Graph Codes for Quantum Error-Correction}
\author{
David J.C. MacKay\\
	Cavendish Laboratory, 
	Cambridge, CB3 0HE. \\
{\tt {mackay@mrao.cam.ac.uk}}
 \and Graeme Mitchison\\
 M.R.C. Laboratory of Molecular Biology, Hills Road, Cambridge, CB2 2QH.\\
{\tt{gjm@mrc-lmb.cam.ac.uk}}
 \and
	Paul L. McFadden\\
 Dept. Applied Maths and Theoretical Physics,
 Cambridge, CB3 0WA \\
{ \tt{p.l.mcfadden@damtp.cam.ac.uk    }}
}
\date{
\ifnum \arabic{longversion} > 0 
 {\tt  
 quant-ph/0304161}
 Version 7.3e\\[0.1in] This is the extended version of a paper
 submitted to {\em IEEE Transactions on Information Theory\/}  May 8, 2003;
 revised January 20th, 2004.
\else
 Submitted to {\em IEEE Transactions on Information Theory\/}  May 8, 2003;
 revised January 20th, 2004.  
 Version 7.3.\\[0.1in] An extended version of this paper is available from {\tt  
 quant-ph/0304161}.
\fi
}

\maketitle
 Indexing terms: Error-correction codes, probabilistic decoding, quantum error-correction.
\begin{abstract}
 We present sparse-graph codes appropriate for
 use in quantum error-correction.

 Quantum  error-correcting codes
 based on sparse graphs are of interest
 for three reasons. First,  the best
 codes currently known for classical channels are based on sparse graphs.
 Second,  sparse-graph codes keep  the number of
 quantum  interactions  associated
 with the quantum  error-correction  process small: a constant number per
 quantum bit, independent of the blocklength.
 Third,  sparse-graph codes  often offer great flexibility with respect to
 blocklength and rate.

 We believe some of the codes we present are unsurpassed by previously
 published  quantum error-correcting codes.
\end{abstract}

\section{Introduction}
 Our aim in this paper is to create {\em useful\/} quantum error-correcting codes.
 To be useful, we think a quantum code must have a large blocklength (since
 quantum computation becomes interesting only when the number of entangled
 qubits is substantial), and it must be able to correct a large number
 of errors. 
 From a theoretical point of view, we would especially like to 
 find, for any rate $R$,
 a family of error-correcting codes with increasing
 blocklength $N$, such that,
 no matter how large $N$ is,
 the number of errors that can be corrected
 is proportional to $N$. (Such codes are called `good' codes.) 
 From a practical point of view, however, we will settle for
 a lesser goal: to be able to make codes with blocklengths in the range $500$--$20,000$ qubits
 and rates in the range 0.1--0.9 
 that can correct the largest number of errors possible.
%
%

 While the existence of `good' quantum 
 error-correcting codes was proved by
\ifnum \arabic{longversion} > 0 
 \citeasnoun{ShorCSS}, their
\else
 Calderbank and Shor \cite{ShorCSS}, their
\fi
 method of proof  was non-constructive.  Recently, a family of 
 asymptotically good quantum codes based on algebraic geometry has been found
 by
\ifnum \arabic{longversion} > 0 
 \citeasnoun{Ashikhmin}
\else
 Ashikhmin {\em et. al.\/} \cite{Ashikhmin}
\fi
 (see also
 \citeasnoun{LingChenXing2001} and \citeasnoun{Matsumoto2002}); however to the best of our 
 knowledge no practical decoding 
 algorithm (\ie, an algorithm for which the decoding time is polynomial in the blocklength) 
 exists for these codes.  Thus, the task of constructing 
 good quantum error-correcting codes for which there exists a practical decoder 
 remains an open challenge.   

 This stands in contrast to the situation for classical error-correction, where
 practically-decodable codes exist which, when optimally decoded, achieve information rates
 close to the Shannon limit.  Low-density parity-check codes
 \cite{Gallager62,Gallager63}
 are an example of such codes.
 A regular low-density parity-check code has
 a parity-check matrix $\bH$ in which each column has 
 a small weight $j$ (\eg, $j=3$) and the weight per row, $k$ is also uniform 
 (\eg, $k=6$).  
 Recently low-density parity-check codes have been shown to have outstanding 
 performance \cite{mncEL,mncN} and modifications to their construction have turned
 them into state-of-the-art codes, both at low rates and large blocklengths 
 \cite{Luby2001b,Richardson2001b,DM_LDPC_ITW98} and at
 high rates and short blocklengths 
 \cite{MacKayHighRate98}. The sparseness of the parity-check matrices
 makes the codes easy
 to encode and decode, even when communicating very close to the Shannon limit. 
 It is worth emphasizing that the sum-product algorithm solves the decoding
 problem for {\ldpcc}s at noise levels  far greater than the maximum
 noise level correctable by any code decoded by a
 traditional bounded-distance decoder.

 This paper explores the conjecture  that the best quantum 
 error-correcting codes will be closely related to the best classical codes.
 By converting classical low-density parity-check codes into 
 quantum codes, we hope to find 
 families of excellent quantum codes.

 Since the parity-check matrix is sparse, a quantum low-density parity-check code
 would have the additional
 attractive property that only a small number of interactions per qubit
 are required in order to determine the error that occurred.
 Moreover, since practical decoding algorithms have been found for classical 
 low-density parity-check codes, it seems likely that a practical decoding 
 algorithm will also exist for  quantum low-density parity-check codes.

 In section \ref{gjmsection} we review the {\sl stabilizer formalism\/}
 for describing quantum error-correcting codes that
 encode a quantum state of  $K$ qubits in $N$ qubits,
 and explain how a general stabilizer code is  related to
 a classical binary code.

%

 In section \ref{djcmsection1} we review sparse graph codes,
 in section \ref{djcmsection2} we discuss dual-containing sparse
 graph codes, and in \secref{subsec.main}
 we present several families of
 classical  sparse graph codes  that satisfy the
 constraints required to make a valid stabilizer code.

 In sections \ref{djcmsection3}  and
 \ref{results} we describe the experimental performance
 of codes from these families on three channels.

 In this paper, we ignore the issue of {\em fault tolerance}: our codes
 correct errors in the encoded quantum state,
 assuming that the encoding and decoding  circuits function perfectly.

\section{Quantum Codes}
\label{gjmsection}
\label{sec.quantumcodes}
 We now review quantum error-correction and the connection
 between quantum codes and classical codes. For further
 reading on quantum codes, we direct the reader to the admirably clear accounts 
 in \cite[Chapter 5]{Lo2001}, \cite[Chapter 7]{Preskill219},
%
%
 and \cite{SteaneCodes}.  Boxes \ref{box1} and \ref{box2} review our notation and conventions.

\newcommand{\ourboxwidth}{6.52in}
\begin{boxfloat}
\begin{framedalgorithmw}{\ourboxwidth}
\small
The Pauli matrices \X, \Y, and \Z\ 
\beq
{\X}=\left(\matrix{ 0 & 1 \cr 1 & 0 \cr}\right),\ \ 
{\Y}=\left(\matrix{ 0 & -i \cr i & 0 \cr}\right),\ \ 
{\Z}=\left(\matrix{ 1 & 0 \cr 0 & \!\!-1 \cr}\right).
\eeq
 have the actions
\eq
\begin{array}{rcl}
 \X (\alpha_0\ket{0}+\alpha_{1}\ket{1}) &=&  \alpha_0\ket{1}+\alpha_{1}\ket{0}    \\
 \Y (\alpha_0\ket{0}+\alpha_{1}\ket{1}) &=&  i(\alpha_0\ket{1}-\alpha_{1}\ket{0}) \\
 \Z (\alpha_0\ket{0}+\alpha_{1}\ket{1}) &=& \alpha_0\ket{0}- \alpha_{1}\ket{1} ;   
\end{array}
\en
thus \X\ is a bit flip,
\Z\ is a phase flip, and \Y\ (ignoring the phase factor $i$) is a combination
of bit and phase flips. \X, \Y\ and \Z\ satisfy
\begin{equation}
{\X}^2={\I} \ \ \ \ {\Y}^2={\I} \ \ \ \ {\Z}^2={\I}
\label{square}
\end{equation}
and
\begin{equation}
\begin{array}{c}
{\X}{\Y}=i{\Z} \\
{\Y}{\Z}=i{\X} \\
{\Z}{\X}=i{\Y} 
\end{array} \:\:\:\:\:\:
\begin{array}{c}
{\Y}{\X}=-i{\Z}\\
{\Z}{\Y}=-i{\X}\\
{\X}{\Z}=-i{\Y}
\end{array}
\label{commute}
\end{equation}
\end{framedalgorithmw}
\caption[a]{The Pauli operators.}
\label{box1}
\end{boxfloat}
\begin{boxfloat}
\begin{framedalgorithmw}{\ourboxwidth}
\small
A Pauli operator on $N$ qubits has the form $c\O_1\O_2 \dots \O_N$,
where each $\O_i$ is one of \I, \X, \Y, or \Z\  and $c=1, -1, i$ or
$-i$. This operator takes $\ket{i_1i_2 \dots i_N}$ to
$c\O_1\ket{i_1}\otimes
\O_2\ket{i_2}\otimes \dots \otimes \O_N\ket{i_N}$. So for instance
${\I\X\Z}(\ket{000}+\ket{111})=\ket{010}-\ket{101}$.
For convenience, we will also sometimes employ a shorthand notation
for representing Pauli operators, in which only the non-identity $\O_i$ 
are given; for example ${\I\X\I\Z\I}$ is
denoted by ${\X}_2{\Z}_4$.
\smallskip

Two Pauli operators {\sl commute\/}
 if and only if there is an even
number of places where they have different Pauli matrices neither of
which is the identity \I. This follows from the relations
(\ref{square}) and (\ref{commute}). Thus for example \X\X\I\ and \I\Y\Z\ do
not commute,  whereas \X\X\I\ and \Z\Y\X\ do commute. If two Pauli operators do not commute, they anticommute, since their individual Pauli matrices either commute or anticommute.
\end{framedalgorithmw}

\caption[a]{Pauli operators on $N$ qubits.
 We reserve the typewriter font (\eg, $\X$)
 for operators that act on a single qubit.
}
\label{box2}
\end{boxfloat}

 The analogue of a classical bit is a {\em qubit}, a quantum state
$\ket{\psi}$ in a two-dimensional complex vector space $H_2$ which can be written
\beq
\ket{\psi}= \alpha_0 \ket{0}+ \alpha_{1} \ket{1}
\eeq
with $\alpha_0$, $\alpha_{1}$ complex numbers satisfying $|\alpha_0|^2+|\alpha_{1}|^2=1$.
 $\ket{\psi}$
is determined up to a phase factor $e^{i\theta}$, so $\ket{\psi}$ and
$e^{i\theta}\ket{\psi}$ define the same state. 

 The quantum state of  $K$ qubits has the form
 $\sum \alpha_{\bs} \ket{\bs}$, where  $\bs$ runs over all binary strings of 
 length $K$, so there are 
 $2^K$  complex coefficients $\alpha_{\bs}$,
 all independent except for the normalization constraint
\eq
 \sum_{\bs=000\ldots00}^{111\ldots11}
	|\alpha_{\bs}|^2                 = 1 .
\en
 For instance,
 $\alpha_{00} \ket{00}+\alpha_{01} \ket{01}+ \alpha_{10} \ket{10}+
 \alpha_{11} \ket{11}$, with
 $|\alpha_{00} |^2+|\alpha_{01} |^2+|\alpha_{10} |^2+|\alpha_{11} |^2=1$
 is the general 2-qubit state (where
 $\ket{00}$ is shorthand for the tensor product $\ket{0} \otimes \ket{0}$).

 Whereas a classical binary $(N,K)$ code protects a {\em
 discrete-valued\/} message $\bs$ taking on one of $2^K$ values by
 encoding it into one of $2^K$ discrete codewords of length $N$ bits,
 quantum error correction has a much tougher job: the
 quantum state of $K$ qubits is specified by $2^K$ {\em
 continuous-valued\/} complex coefficients $\alpha_{\bs}$, and the aim
 is to encode the state into a quantum state of $N$ qubits in such
 a way that errors can be detected and corrected, and all $2^K$
 complex coefficients perfectly restored, up to a phase factor.

 The errors that must be corrected include continuous errors
 corresponding to unitary rotations in the quantum state space, and
 errors in which an accidental `measurement' causes the quantum state
 to collapse down into a subspace.  As an example of a continuous
 error, if one qubit is physically embodied in the spin of a
 particle,  an environmental magnetic field might induce a rotation
 of the spin through some arbitrary angle. This rotation of the spin
 corresponds to a unitary transformation of the quantum state
 vector. It might seem impossible to correct such errors,
 but it is one of the triumphs of quantum information theory that,
 when the state is suitably encoded, error correction is possible.

Consider the following encoding of a single qubit in three qubits:
\eq
\begin{array}{rcl}
\ket{\bar 0} &=& {1 \over \sqrt{2}}(\ket{000}+\ket{111})\\
\ket{\bar 1} &=& {1 \over \sqrt{2}}(\ket{000}-\ket{111}),
\end{array}
\en
(where the bar denotes the encoded state).  A general state of one
 qubit, $\alpha_{0} \ket{0} + \alpha_{1} \ket{1}$, is encoded into
 $\alpha_{0} \ket{\bar 0} + \alpha_{1} \ket{\bar 1}$.  We will show
 that the receiver can detect and correct single qubit flips if he
 measures two diagnostic operators \Z\Z\I\ and \I\Z\Z, which we call
 the quantum syndrome.  (These two observables commute, so they can
 both be measured simultaneously.)  Imagine that the first qubit
 undergoes a flip $\ket{0} \leftrightarrow \ket{1}$, so $\alpha_{0}
 \ket{\bar 0} + \alpha_{1} \ket{\bar 1}$ becomes $\alpha_{0} {1 \over
 \sqrt{2}}(\ket{100}+\ket{011}) + \alpha_{1} {1\over
 \sqrt{2}}(\ket{100}-\ket{011})$.  Now, measuring \Z\Z\I\ gives $-1$,
 and \I\Z\Z\ gives $+1$, for any value of the coefficients
 $\alpha_{0}$ and $\alpha_{1}$. The four possible outcomes of $\pm 1$
 for the two operators correspond to the three possible bit-flips and
 to no flip. Thus measuring the observables \Z\Z\I\ and \I\Z\Z\ gives
 an error diagnosis analogous to the classical syndrome. As in the
 classical case, the appropriate correction can be made after syndrome
 measurement. In the above example, one applies the bit-flip operator
 $\X$ to the first qubit, thus restoring the state to $\alpha_{0}
 \ket{\bar 0} + \alpha_{1} \ket{\bar 1}$.

 Now consider an error that lies in the continuum of rotations: the
 operator $( \cos \theta ) {\I} + i ( \sin \theta ) {\X}$ applied to
 the first qubit in the encoded state. For simplicity, we assume the
 encoded state is $\ket{\bar 0}$; the error operator takes $\ket{\bar
 0}$ to
\eq 
{1 \over \sqrt{2}}
 \left[
   \left(
        \cos \theta\ket{0}+i \sin\theta\ket{1}
   \right)
		\ket{00}+
 \left(i
\sin \theta\ket{0}+\cos\theta\ket{1}
 \right)
		\ket{11}
\right].
\en 
Measuring the syndrome operators {\tt{ZZI}} and {\tt{IZZ}}
causes the state to collapse onto the
original encoded state $\ket{\bar 0}$ with probability $\cos ^2
\theta$ or onto the bit-flipped state $\ket{100}+\ket{011}$ with probability $\sin^2
\theta$. In the former case we get the syndrome ($+1$,$+1$), and in the
latter ($-1$,$+1$), so the appropriate correction can be
applied. Exactly the same collapse-and-correct procedure works if one
has a general encoded qubit $\alpha_{0} \ket{\bar 0} + \alpha_{1}
\ket{\bar 1}$.
%
 Thus, for a cunning choice of syndrome measurement,
 collapse in effect `picks' a discrete error, and allows one to
 think of bit flips much as one does in classical
 error-correction.

However, bit flips are not the only types of error that can occur in
quantum states. In a `phase flip', which corresponds to the application
of \Z\ to a qubit, the coefficient of $\ket{0}$ remains unchanged but
the sign of the coefficient of $\ket{1}$ is switched. In the above
example, the encoded state $\ket{\bar 0}$ would be taken to $\ket{\bar
1}$ (and vice versa) by a phase flip of any of its qubits:
\eq
 \Z_n \ket{\bar 0} = \frac{1}{\sqrt{2}}
  \Z_n ( \ket{000} + \ket{111}) = \frac{1}{\sqrt{2}} ( \ket{000} - \ket{111})
 = \ket{\bar 1} \:\:\:\mbox{for $n=1$, 2, or 3},
\en
 and this error could not be
 corrected since the corrupted state is a codeword.

 The first quantum code able to correct both bit and phase flips was
 discovered by
\ifnum \arabic{longversion} > 0 
 \citeasnoun{Shor},
\else
 Shor \cite{Shor},
\fi
 and it encodes a single qubit in nine
 qubits:
\begin{eqnarray}
\ket{\bar 0} &=& {1 \over \sqrt{8} }(\ket{000}+\ket{111})(\ket{000}+\ket{111})(\ket{000}+\ket{111})\nonumber \\
\ket{\bar 1} &=& {1 \over \sqrt{8} }(\ket{000}-\ket{111})(\ket{000}-\ket{111})(\ket{000}-\ket{111}),
\label{shorcode}
\end{eqnarray}
 Here, a single bit flip in the first three qubits can be detected by
 measuring ${\Z}_1{\Z}_2$, ${\Z}_2{\Z}_3$ (as in the
 previous code we considered), and similarly ${\Z}_4{\Z}_5$,
 ${\Z}_5{\Z}_6$, ${\Z}_7{\Z}_8$ and
${\Z}_8{\Z}_9$ detect a bit flip in the remaining qubits. But
one can also detect phase flips, by measuring {\operatorfont{XXXXXXIII}} and
{\operatorfont{IIIXXXXXX}}, and these two diagnostic operators together with the six
\Z-containing ones above form a commuting set which can therefore be
measured simultaneously. Note that the outcomes of the \X\
measurements do not determine which of the qubits in each block of
three underwent a phase flip, but this knowledge is unnecessary for
correcting the state, because changing the sign of a particular qubit
in the block corrects a sign change of any one of the three. Note also
that a combined bit and phase flip on the same qubit, a \Y\ error, can
be detected and corrected. For instance, a \Y\ error on the first
qubit gives a $-1$ on measuring ${\Z}_1{\Z}_2$ and
{\operatorfont{XXXXXXIII}}, and $+1$ for all the other operators.

As any unitary transformation on one qubit can be written as a
weighted sum of \I, \X, \Y, and \Z\  (because these
four operators span the space of $2 \times 2$ matrices), it follows
that any error of this unitary type on one qubit can be corrected.
Furthermore, more general errors, measurements for instance, can be
represented as sums of these operators.

Suppose, for example, an interaction with the environment has the
effect of a measurement of the observable ${\operatorfont M}=(\cos \theta) {\Z}
+ (\sin
\theta) {\X}$ on the first qubit. Let
 the projections corresponding to the eigenvalues $+1$ and $-1$ be $\tP$ and $\tQ$,
respectively. Then $\tP+\tQ={\I}$ and $\tP-\tQ={\operatorfont M}$,
so the projection onto the $+1$ eigenspace is $({\I+ \tM})/2$.
Thus, if the measurement outcome (which is not known to us) is +1,
 the projection takes an encoded state $\ket{\psi}$ to the state
\eq 
\ket{\tilde \psi}={\cal N}({\I}_1+\cos \theta {\Z}_1 + \sin \theta
{\X}_1)\ket{\psi},
\en
where ${\cal N}$ is a normalization constant. When all eight
diagnostic operators are measured, the probabilities of the various
syndromes are as shown in  \tabref{star}.
Having diagnosed the error that is present after syndrome measurement,
it can be corrected, despite the fact that the original error in this
case was not even a unitary transformation.

 The foregoing examples all involve an error on one qubit only. An
 error operator that affects several qubits can be written as a weighted sum
 $\sum c_\alpha P_\alpha$ of Pauli operators acting on those qubits and
 acting as the identity on other qubits. (A general operator on $k$
 qubits can be written as such a sum because the $4^k$ possible Pauli
 operators span the space of $2^k \times 2^k$ complex matrices.)  Given
 such an error and a codeword $\ket{\psi}$, a suitable set of
 diagnostic operators will `pick' an individual error term $P_{\tilde
 \alpha}\ket{\psi}$ from the sum, and an error correction procedure applied to this
 term will restore the original codeword.

 An error may  act not only on the code qubits but  on the
 environment too. Given an initial state that is a product $\ket{\psi}
 \ket{\phi}^e$ of code qubits and environmental variables, any error acting on
 both the code and the environment can be written as a weighted sum
 $\sum c_{\alpha,\beta} P_\alpha P_\beta^e$ of Pauli operators that act
 on code and environment qubits.
 If $S_i$ are diagnostic operators that pick the error term $P_{\tilde
 \alpha}\ket{\psi}$, then the operators $S_i I^e$ will pick terms $\sum_{\tilde
 \beta}
 c_{\tilde\alpha,\tilde \beta}P_{\tilde \alpha} \ket{\psi} P_{\tilde
 \beta}^e \ket{\phi}^e$ from the corrupted state, and these  terms can 
 be written as $(P_{\tilde \alpha} \ket{\psi} )\ket{\mu}^e$ for some new
 environmental state $\ket{\mu}^e$. 
 So
 measuring the syndrome restores a product state of qubits and
 environment, and in this sense the code and the environment evolve
 independently and we may ignore the environment in what follows.

\begin{table}
\begin{center}
\begin{tabular}{ccccc}
\toprule
    & Error: & $\I_1$ & ${\operatorfont X}_1$ & ${\operatorfont Z}_1$ \\
\midrule 
 & Probability:   &${1 \over 2}$ & ${1 \over 2}\sin^2 \theta$ & ${1 \over 2}\cos^2 \theta$ \\
\midrule
Stabilizer &&& Syndrome &\\
\midrule
${\operatorfont ZZIIIIIII}$ && +1 & $-1$ & +1\\
${\operatorfont IZZIIIIII}$ && +1 & +1 & +1\\
${\operatorfont IIIZZIIII}$ && +1 & +1 & +1\\
${\operatorfont IIIIZZIII}$ && +1 & +1 & +1\\
${\operatorfont IIIIIIZZI}$ && +1 & +1 & +1\\
${\operatorfont IIIIIIIZZ}$ && +1 & +1 & +1\\
${\operatorfont XXXXXXIII}$ && +1 & +1 & $-1$\\
${\operatorfont IIIXXXXXX}$ && +1 & +1 & +1\\
\bottomrule
\end{tabular}
\end{center}
\caption{The first column shows a set of stabilizers for the Shor code, and the three right-hand columns show syndromes
 corresponding to three particular error operators, together with their probabilities.
\label{star}}
\end{table}

\subsection{Stabilizer codes}

The scene is now set to define a general way of making quantum codes:
the stabilizer framework. A stabilizer is essentially what we have
been calling a `diagnostic operator'. We now begin with a set of such
operators and use it to define a code.

A stabilizer group  ${\cal S}$
consists of a set of Pauli operators on $N$
qubits closed under multiplication, with the property that any two
operators in the set commute, so that all can be measured
 simultaneously. It is enough to check this
commutation property on a set of
generators of ${\cal S}$, i.e., on a set $\{ S_i \}$ that generate all of
${\cal S}$ under multiplication. For instance, we could take the
diagnostic operators associated with the Shor code (see  \tabref{star})
%
%
as generators for a set of stabilizers ${\cal S}$. The full set ${\cal
S}$ will include operators like {\operatorfont{ZIZIIIIII}} (the product of the first
two generators) and {\operatorfont{YYXXXXIII}} (the product of the first and seventh),
and so on.

Given a set of stabilizers, a codeword is defined to be a state
$\ket{\psi}$ that is a +1 eigenstate of all the stabilizers, so
\begin{equation}
\S_i\ket{\psi}=\ket{\psi} \mbox{  for all $i$}.
\label{codeword}
\end{equation}
Consider what the codewords of the stabilizers shown in 
\tabref{star} must be. If $\ket{\psi}=\sum_\bs \alpha_\bs \ket{\bs}$,
\eqref{codeword} applied to the first two stabilizers in the \tablenoun\ 
implies that the first three bits in any binary string $\bs$ in the
sum must be $000$ or $111$, and the same is true for the two other
groups of three. From the last two stabilizers in the \tablenoun, we deduce
that strings $\bs$ with an odd number of 1s all have equal
coefficients $\alpha_\bs$, and similarly for those with an even number
of 1s. It follows that the code is generated by $\ket{\bar 0}$ and
$\ket{\bar 1}$ given by
\eqref{shorcode}. Thus we recover the original Shor code.

Consider now a set of error operators $\{ \E_\alpha\}$, i.e. Pauli
operators taking a state $\ket{\psi}$ to the corrupted state
$\E_\alpha \ket{\psi}$. A given operator $\E_\alpha$ either commutes
or anticommutes with each stabilizer generator $S_i$ (see Box
\ref{box2}). If $\E_\alpha$ commutes with $S_i$ then
\eq 
\S_i\E_\alpha\ket{\psi}=\E_\alpha\S_i \ket{\psi}=\E_\alpha\ket{\psi},
\en 
so $\E_\alpha \ket{\psi}$ is a $+1$ eigenstate of $\S_i$. Similarly,
if it anticommutes, $\E_\alpha \ket{\psi}$ is a $-1$ eigenstate of
$\S_i$. Thus $\E_\alpha \ket{\psi}$ is an eigenstate of the joint
measurement of all the stabilizer generators, and the outcome of this
measurement -- the syndrome -- is completely determined by the
commutation properties of $\E_\alpha$ with the stabilizers.  Thus the
syndrome is determined by the error operator and is independent of the
state $\ket{\psi}$, which implies that we learn nothing about the
state in measuring the syndrome. This is important since, in quantum
mechanics, a state is usually damaged when a measurement yields
information about it.

A sufficient condition for
 the set of error operators $\{ \E_\alpha\}$
 to be
  {\em correctable\/} is
   that
each operator of the set
should have a distinct syndrome, so the syndrome
determines the index $\alpha$. Correction can then be performed by
applying the specific $\E_\alpha$ to the corrupted state, since
$\E_\alpha^2$ is the identity operator up to some phase factor.
 A  set of error operators $\{ \E_\alpha\}$
  is also 
      correctable
       if any two operators $\E_\alpha$ and
$\E_\beta$
that
have the same syndrome
 differ by a
stabilizer. Thus $\E^\dagger_\alpha \E_\beta$ is a stabilizer, $\S$
say, so $\E_\beta=\E_\alpha \S$. Then $E_\beta \ket{\psi}=E_\alpha S
\ket{\psi}=E_\alpha \ket{\psi}$, so $E_\alpha$ and $E_\beta$
generate the same corrupted state and can therefore be corrected by
the same operator.

An error arising from any linear combination of operators from a
correctable set of Pauli operators $\{ \E_\alpha\}$ can be
corrected. Just as in the example of the Shor code, syndrome
measurement collapses the state onto one of the syndrome eigenspaces,
and the original state can then be restored.

\ifnum \arabic{longversion} > 0 

 Most realistic error mechanisms will generate linear combinations that
 include uncorrectable error operators. For instance, consider an error
 process ${\cal E}_{\tU}$ in which each qubit experiences an independent
 perturbation, undergoing a unitary transformation
\beq
\tU=(1-p^2-q^2-r^2)^{1/2}\I+ip\X+iq\Y+ir\Z,
\eeq
 where $p$, $q$ and $r$ are small real numbers. Starting from an
 encoded state $\ket{\psi}$, the perturbed state
 includes
 terms $c_\alpha E_\alpha \ket{\psi}$ for every possible Pauli operator
 $E_\alpha$. There will be many uncorrectable operators amongst them,
 but if the code is well-chosen for the given error model,
 the coefficients $c_\alpha$ will be
 small for these particular operators.

Suppose that $p$, $q$ and $r$ are chosen from a distribution $P(x)$
that is symmetric about zero, so $P(x)=P(-x)$. If $\rho$ is the
density matrix for the state of a single qubit, its state
$\tilde \rho$ after $\tU$ is applied is
\beq
\tilde \rho=\int P(p)P(q)P(r) \tU  \rho \tU^\dagger dp \, dq \, dr.
\eeq
The symmetry of $P$ ensures that cross-terms like $\X\rho\Y$ are zero,
so
\beq
\tilde \rho =(1-3u)\rho+u \left[\X\rho\X+\Y\rho\Y+\Z\rho\Z \right],
\label{eq.depol}
\eeq
where $u=\int x^2 P(x)\, dx$.  This noise model is
 known  as the depolarizing channel.

Thus the error caused when $\tU$, with randomly chosen $p$, $q$ and $r$,
is applied to a qubit is indistinguishable, by any quantum mechanical
measurement, from a random process that leaves the state unchanged
with probability $1-3u$ or applies $\X$, $\Y$ or $\Z$ with probability
$u$. 

It follows that the error process ${\cal E}_{\tU}$, which applies $\tU$
independently to each qubit, is indistinguishable from a process that
applies one of $\X$, $\Y$ or $\Z$ with probability $u$ to each qubit,
despite the very different characters of the two underlying
mechanisms. We can either think of errors being caused by the presence
of uncorrectable terms $c_\alpha E_\alpha \ket{\psi}$ when each qubit
is given a small perturbation by $\tU$, or by the chance occurrence of
an uncorrectable combination of bit or phase flips. The latter process
closely resembles a classical bit-flip error process.

We now make the
analogy between quantum and classical codes more precise.
\else
\fi

\subsection{The relationship between quantum and classical codes}
\label{sec2.2}
Given any Pauli operator on $N$ qubits, we can write it uniquely as a
product of an \X-containing operator and a \Z-containing operator and
a phase factor (+1, $-1$, $i$, or $-i$). For instance,
\beq
{\operatorfont{XIYZYI}}=  \begin{array}{r@{}l}
                          - ({\operatorfont{XIXIXI}}) & \times \\
 ({\operatorfont{IIZZZI}})                            & . \end{array}  
\eeq
We now express the \X\ operator as a binary string of length $N$, with
`1' standing for \X\ and `0' for I, and do the same for the \Z\
operator. Thus each stabilizer can be written as the \X\ string followed
by the \Z\ string, giving a matrix of width $2N$. We mark the boundary
between the two types of strings with vertical bars, so, for instance,
the set of generators in \tabref{star} appears as the matrix $\MM$, where
\eq 
\MM =\left(\begin{array}{*{9}{c}|*{9}{c}}&&&&\X&&&&& &&&&\Z&&&\\
1&1&1&1&1&1&0&0&0& 0&0&0&0&0&0&0&0&0\\
0&0&0&1&1&1&1&1&1& 0&0&0&0&0&0&0&0&0\\
0&0&0&0&0&0&0&0&0& 1&1&0&0&0&0&0&0&0\\ 
0&0&0&0&0&0&0&0&0& 0&1&1&0&0&0&0&0&0\\
0&0&0&0&0&0&0&0&0& 0&0&0&1&1&0&0&0&0\\
0&0&0&0&0&0&0&0&0& 0&0&0&0&1&1&0&0&0\\
0&0&0&0&0&0&0&0&0& 0&0&0&0&0&0&1&1&0\\
0&0&0&0&0&0&0&0&0& 0&0&0&0&0&0&0&1&1
\end{array} \right).
\en 

The commutativity of stabilizers now appears as orthogonality of rows
with respect to a {\em twisted product\/}
 (also known as a symplectic product): if row $m$ is $r_m=(x_m,z_m)$, where $x_m$ is
the \X\ binary string and $z_m$ the \Z\ string, then the twisted
product $\odot$ of rows $m$ and $m'$ is
\eq 
r_m \odot  r_{m'}=x_m\cdot z_{m'} + x_{m'} \cdot z_m \ \bmod 2,
\en 
where `$\cdot$' is the usual dot product, $x_m\cdot z_{m'}=\sum_i
x_{mi}z_{m'i}$. The twisted product is zero if and only if there is an
even number of places where the operators corresponding to rows $m$
and $m'$ differ (and are neither the identity), i.e., if the operators
commute. If we write $\MM$ as $\MM=(\BBB|\AAA)$, then the condition $r_m
\odot r_{m'}=0$ for all $m$ and $m'$ can be written compactly as
\eq
\BBB\AAA^{\T}+\AAA\BBB^{\T}= {\bf 0}.
\label{twistzero}
\en

A Pauli error operator $\E$ can be interpreted as a binary string
$\be$ of length $2N$. Our convention is that we reverse the order of
the \X\ and \Z\ strings in the error operator, so, for instance the
binary string
\[
\begin{array}{c|c}
10000001 & 010000001
\end{array}
\]
(with a `$|$' inserted for interpretational convenience), corresponds
to the operator ${\Z}_1{\X}_2{\Y}_8$. With this convention, the
ordinary dot product (mod 2) of $\be$ with a row of the matrix is zero
if $\E$ and the stabilizer for that row commute, and 1 otherwise. Thus
the quantum syndrome for the noise is exactly the classical syndrome
$\MM \be$, regarding $\MM$ as a parity check matrix and $\be$ as binary
noise.

We can now define the conditions for error correction in this
classical setting. If there is a set of binary noise vectors $\{
\be_\alpha\}$ that have distinct syndromes, then the errors are
correctable, and so are the corresponding errors in the quantum
code. However, we also know the errors are correctable under the
relaxed requirement that any operators $E_{\alpha}$ and $E_{\beta}$
with the same syndrome differ by a stabilizer. Any stabilizer
$S_\alpha$ in ${\cal S}$ can be written as a product of some subset of
the generator set $\{S_i\}$, and $S_\alpha$ is equivalent to a binary
string generated by adding rows of $\MM$, in other words to an element
of the dual code generated by $\MM$.

This more lax requirement applies to the Shor code shown above. The
binary strings
\[
\begin{array}{c|c}
100000000 & 000000000\\
010000000 & 000000000
\end{array}
\]
have the same syndrome, but they differ by an element of the dual
code, namely the stabilizer ${\Z}_1{\Z}_2$. 

In conclusion, the properties of stabilizer codes can be inferred from
those of a special class of classical codes. Given any binary matrix
of size $M_Q \times 2N$ that has the property that the twisted product
of any two rows is zero, an equivalent quantum code can be constructed
that encodes $N-M_Q$ qubits in $N$ qubits. If there is a set of errors
for the classical code that are uniquely characterized by their
syndromes, or differ by an element of the dual code if they have the
same syndrome, then the corresponding error operators in the quantum
code can be corrected.

\ifnum \arabic{longversion} > 0 
\subsection{Examples of stabilizer codes}
\subsubsection{Cyclic codes}
 An elegant $(N,K)=(5,1)$ quantum code is
 generated by the  four stabilizers given below alongside their equivalent matrix:
\beq
\begin{array}{c} \mbox{Stabilizers} \\
\operatorfont  XZZXI \\
\operatorfont  IXZZX \\
\operatorfont  XIXZZ \\
\operatorfont  ZXIXZ 
\end{array}
 \:\:\:\:
 \:\:\:\:
 \:\:\:\:
 \:\:\:\:
\MM =\left(\begin{array}{ccccc|ccccc}
    &&\X&&&&&\Z&&\\
1&0&0&1&0 & 0&1&1&0&0 \\
0&1&0&0&1 & 0&0&1&1&0 \\
1&0&1&0&0 & 0&0&0&1&1 \\
0&1&0&1&0 & 1&0&0&0&1 
\end{array}
 \right).
\eeq

All the twisted product of rows in $\MM$ vanish, so $\MM$ defines a
commuting set of stabilizers and hence a quantum code. A correctable
set of errors consists of all operators with one non-identity term, \eg\
${\operatorfont{XIIII}}$, ${\operatorfont{IIIYI}}$ or
${\operatorfont{IZIII}}$. These correspond to binary strings such as
$00000|10000$ for ${\operatorfont{XIIII}}$, $00010|00010$ for
${\operatorfont{IIIYI}}$, and so on. There are 15 of these, and each
has a distinct syndrome, thereby using up all the $2^4-1$ possible
non-zero syndromes. This is therefore a perfect quantum code.

If one adds a fifth row $00101|11000$ to $\MM$, so that the
$\X$ and $\Z$ submatrices are cyclic, the fifth row is redundant,
being the sum of the other four rows (its stabilizer is the product of
the other four). 


\subsubsection{CSS codes}

\else
\subsection{CSS codes}
\fi

An important class of codes, invented by Calderbank, Shor \& Steane
\cite{ShorCSS,SteaneCSS}, has the form
\eq 
\MM=\left(\begin{array}{c|c} \AA& {\bf 0}\\
{\bf 0} & \BB \end{array} \right),
\en 
where $\AA$ and $\BB$ are $M \times N$ matrices. Requiring $\AA
\BB^{\T}={\bf 0}$ ensures that (\ref{twistzero}) is satisfied. As there are $M_Q=2M$ stabilizer conditions applying to $N$ qubit states, $N-2M$ qubits are encoded in $N$ qubits.

\ifnum \arabic{longversion} > 0 
\subsubsection{CSS codes based on dual-containing codes.}
\else
\subsubsection{CSS codes based on dual-containing codes.}
\fi

If $\AA=\BB$, $\MM$ has the particularly simple form
\eq 
\MM=\left(\begin{array}{c|c} \AA&  {\bf 0}\\
{\bf 0}& \AA
\end{array} \right).
\label{eq.dualcon}
\en 
\Eqref{twistzero} is satisfied if $\AA \AA^{\T}={\bf 0}$. This is
equivalent to ${\cal C}^\perp(\AA) \subset {\cal C}(\AA)$, where
${\cal C}(\AA)$ is the code having $\AA$ as its parity check matrix
and ${\cal C}(\AA)^\perp$ is its dual code. We call such a code
a `dual-containing' code; it is the type of code we are most
concerned with in this paper. (Dual-containing codes are
 also known as `weakly self-dual codes'.)

Any state of the form
\eq
\ket{\psi}=\sum_{\bx \in {\cal C}^\perp(\AA)} \ket{\bx+\by},
\label{encodedstate}
\en
with $\by
\in {\cal C}(\AA)$ and `$+$' denoting addition mod 2, is a codeword. This is because each \X-stabilizer $S$ permutes
the terms in the sum, so $S\ket{\psi}=\ket{\psi}$, and the
\Z-stabilizers leave each term in the sum unchanged, since the
dual-containing property implies that $\bx+\by$ is a codeword of $\cal
C(\AA)$ if $\bx \in {\cal C}^\perp(\AA)$ and $\by \in {\cal C}(\AA)$.
The most general codeword has the form
\beq
\ket{\psi} = \sum_{\by \in {\cal C}(\AA)} \alpha_{\by} \sum_{\bx \in {\cal C}^\perp(\AA)} \ket{ \bx + \by }.
\label{general}
\eeq

An example of a dual-containing code is Steane's 7 qubit code, defined
by the Hamming code
\eq 
\AA=\left(\matrix{0&0&0&1&1&1&1\cr
0&1&1&0&0&1&1\cr
1&0&1&0&1&0&1\cr}\right).
\en 
The rows have an even number of 1s, and any two of them overlap by an
even number of 1s, so ${\cal C}^\perp(\AA) \subset {\cal C}(\AA)$. Here
$M=3$, $N=7$, so $N-2M=1$, and thus 1 qubit is encoded in 7 qubits.


\ifnum \arabic{longversion} > 0 
\subsubsection{Codes over GF(4)}

Let the elements of GF(4) be 0, 1, $w$ and $w^2=\bar w=1+w$. One can write a row
$r_1=a_1a_2 \dots a_n|b_1b_2 \dots b_n$ in the binary string
representation as a vector over GF(4), $\rho_1=(a_1+b_1w, a_2+b_2w,
\ldots, a_n+b_nw)$. Given a second row $r_2=c_1c_2 \dots c_n|d_1d_2 \dots d_n$, with $\rho_2= (c_1+d_1w, c_2+d_2w, \ldots, c_n+d_nw)$, the Hermitian inner product is defined by
\eq
\rho_1 \cdot \rho_2=\sum_i (a_i+b_i \bar w)(c_i+d_i w)=\sum_i
\left[ (a_ic_i+b_id_i+b_ic_i)+(a_id_i+b_ic_i)w\right].
\label{herm}
\en
Now the coefficient of $w$ on the right-hand side of this equation is
\beq
(a_id_i+b_ic_i)= r_1 \odot r_2 
\eeq
so if the Hermitian inner product of two rows is zero then $r_1 \odot
r_2=0$ (though the converse does not hold).  Thus a code over GF(4)
that contains its Hermitian dual defines a stabilizer code, since we
can interpret its rows as binary strings whose twisted products
vanish \cite{GF4codes}. However, not all stabilizer codes can be obtained from GF(4)
codes in this way, since, as we have just noted, $r_1 \odot r_2=0$
does not imply that the Hermitian product $\rho_1 \cdot \rho_2$ is zero.

\subsection{A quantum circuit for decoding}
\begin{figure}
\centerline{\epsfbox{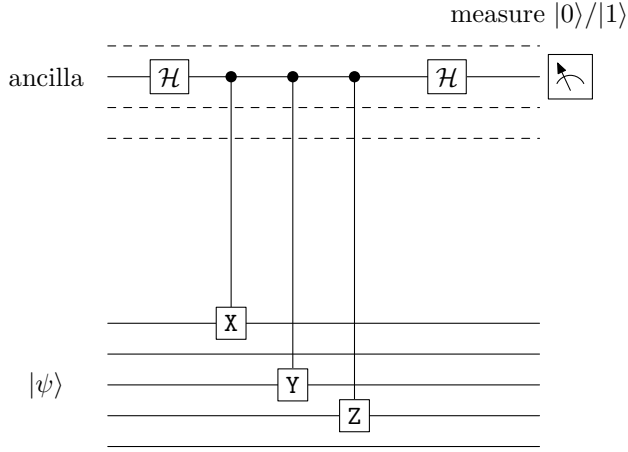}}
\caption{Measuring the syndrome of a quantum code.
 There is an ancilla for each stabilizer.
 Here the operations for an ancilla corresponding to the stabilizer $S_\alpha={\operatorfont{XIYZI}}$ are shown. The black dot on the ancilla's line indicates that the ancilla controls the operator (e.g. ${\operatorfont{X}}$) in the attached box acting on the state $\ket{\psi}$ to be decoded. The boxes labelled `${\cal H}$' carry out Hadamard transforms on the ancilla.
\label{decode}}
\end{figure}

We consider decoding first as this can be carried out by circuits that
are quite simple. In the circuit shown in  \figref{decode}, an
extra line, corresponding to an additional ancillary qubit, controls a
set of operations on the encoded qubits. Here, `control' means that
when the ancilla is $\ket{0}$, no operation is performed on the
encoded qubits, and when the ancilla is $\ket{1}$ the operation shown
in the box (${\tt X}$, ${\tt Y}$ or ${\tt Z}$) is carried out. Taken
together, the individual controlled operations correspond to a
controlled stabilizer operator $S^c_\alpha$ on N qubits, and in the
figure the stabilizer  $S_\alpha$ is assumed to be ${\operatorfont{XIYZI}}$.

An initial and final Hadamard operation 
\beq
{\cal H}={1 \over \sqrt 2}\left(\matrix{1&1\cr 1&-1\cr}\right)
\eeq
is carried out on the ancilla. The effect of this is that a final
measurement of the ancilla in a $\ket{0}/\ket{1}$ basis gives outcome
0 when $S_\alpha$ has outcome $+1$, and 1 when $S_\alpha$ has outcome
$-1$. This follows because
\begin{eqnarray}
(\I \otimes {\cal H}) S^c_\alpha(\I \otimes {\cal H}) \ket{\psi}\ket{0}
 &=& (\I \otimes {\cal H}) S^c_\alpha \ket{\psi}(\ket{0}+\ket{1})/\sqrt 2 \\
&=& (\I \otimes {\cal H}) (\ket{\psi} \ket{0}+S_\alpha \ket{\psi}\ket{1})/2\\
&=& \left[\ket{\psi} (\ket{0}+\ket{1})+S_\alpha \ket{\psi}(\ket{0}-\ket{1})\right]/2\\
&=& {1 \over 2} ({\operatorfont{I}}+S_\alpha) \ket{\psi}\ket{0}+{1 \over 2} ({\operatorfont{I}}-S_\alpha) \ket{\psi}\ket{1},
\end{eqnarray}
and thus measuring the ancilla projects $\ket{\psi}$ onto the
eigenstates of $S_\alpha$.

\begin{figure}
\centerline{\epsfbox{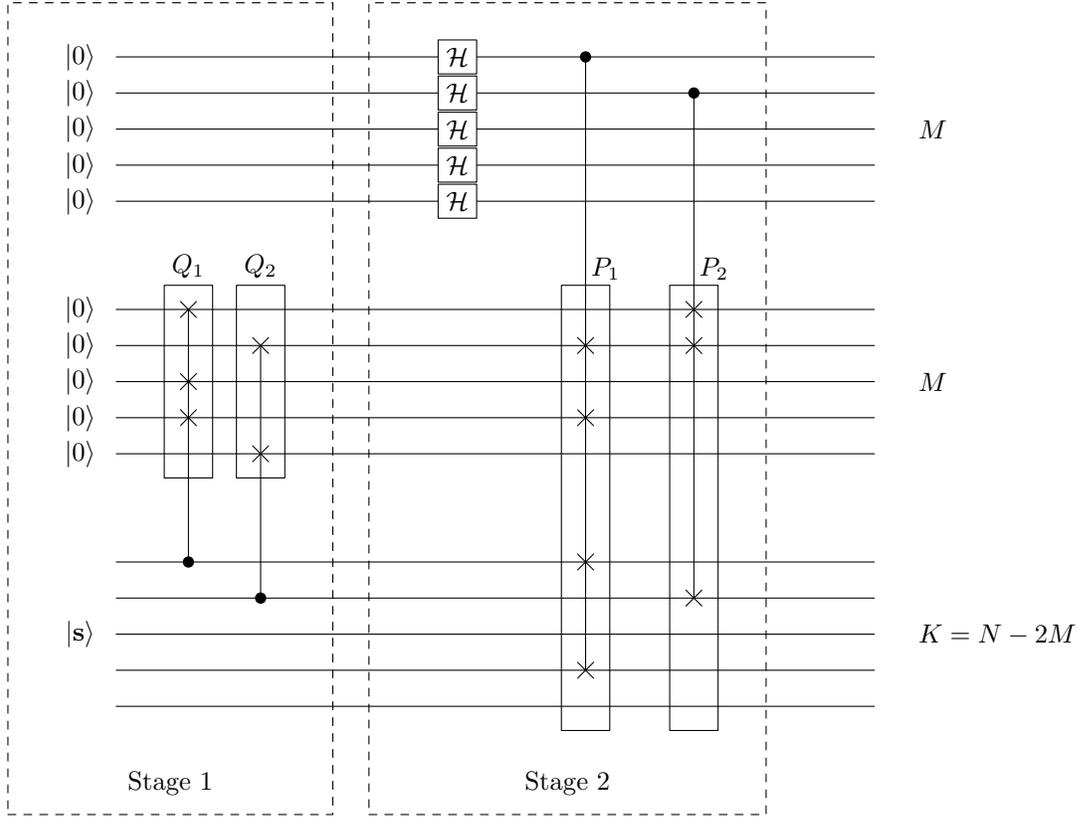}}
\caption{Encoding a quantum state.
 The first stage
 applies a series of controlled operations that carry out
 multiplication by the matrix $\QQ$. This means that the $k$th box,
 $Q_k$, controlled by the $k$th qubit in the last $\K = N-2M$ qubits, has a
 cNOT at the $m$th qubit within the box if and only if the $k$th
 column of $\QQ$ has a 1 in its $m$th position, i.e.,
 $Q_{mk}=1$. The second stage adds the {\em rows} of $\PP$ to the
 last $N-M$ qubits by applying cNOT operations at all positions where
 the $m$th row of $\PP$ has a 1. Thus the box labelled $\P_m$ applies
 cNOTs where the $m$th row of $\PP$ has a 1, and is controlled by the
 $m$th qubit in the top block of $M$ qubits.
\label{encode}} 
\end{figure}
\subsection{A quantum circuit for encoding}
\label{sec.qencode}

We show now how to encode, in the case of a dual-containing
code defined by a full rank matrix $\AA$, with $N > 2M$. We call the
 number of source qubits  $K=N-2M$. First, some algebra: 

A set of row operations plus reordering of columns allows $\AA$ to be
transformed to $\tilde \AA=[\bf I,\PP]$, where the notation indicates
that the $M \times M$ identity matrix $\bf I$ and an $M \times (N-M)$
binary matrix $\PP$ are placed side by side to form the $M \times N$
matrix $\tilde \AA$. We can assume that the reordering of columns was
not necessary (in other words, we chose the right ordering of columns
in $\AA$ to begin with). Now, $\PP$ has full rank, because if not,
then row operations on $\tilde \AA$ could create a row ${\bf r}$ that
was zero in the last $N-M$ bits but had some non-zero bits in the
first $M$. Since ${\bf r} \in {\cal C}^\perp(\tilde \AA)={\cal
C}^\perp(\AA)$ and ${\cal C}^\perp(\AA) \subset {\cal C}(\AA)$, this
must be a codeword of ${\cal C}(\AA)$ and hence of ${\cal C}(\tilde
\AA)$. But the inner product of some rows of $\tilde \AA$ with this
codeword must be non-zero, having a 1 bit overlap in the first $M$
bits and no overlap elswhere, and this is a contradiction.

Thus $\PP$ has full rank, and we can apply row operations to $\PP$ in
turn to obtain $\tilde \PP=[{\bf I},\QQ]$ for some $M \times K$
matrix $\QQ$. For any length-$K$
 binary string $\bff$, $[\QQ \bff,\bff]$ is a
codeword of ${\cal C}(\tilde \PP)$, hence of ${\cal C}(\PP)$, and so
$[{\bf 0},\QQ \bff,\bff]$ is a codeword of ${\cal C}(\AA)$. Furthermore, if
two $\bff$s are distinct, the resulting codewords cannot differ by an
element of ${\cal C}^\perp(\AA)$, since any non-zero element of ${\cal
C}^\perp(\AA)$ has 1s in the first $M$ bits whereas the codewords
obtained by our construction have zeros in the first $M$ bits. Thus
each $\bff$ gives a unique equivalence class in ${\cal C}$ relative to
the subspace ${\cal C}^\perp$.

We can use this construction to encode $K = N-2M$ qubits in $N$ qubits by
the circuit shown in  \figref{encode}. Given a $K$ qubit state
$\ket{\bs}_{\K}$, where $\bs$ is a $\K$ bit string, the first stage
of the circuit
carries out the transformation
\beq
\ket{\bf 0}_M\ket{\bf 0}_M\ket{\bs}_{\K} \rightarrow \ket{\bf 0}_M\ket{\QQ \bs}_M\ket{\bs}_{\K}
\eeq
using the cNOT (controlled NOT, or controlled \X) operations shown in
the boxes $Q_1$, $Q_2$, etc..
The next stage of the circuit (`stage 2')
first applies a Hadamard transform on the first $M$ qubits, taking the
initial $\ket{\bf 0}_M$ state to
\beq
\ket{\bf 0}_M\rightarrow \prod^M {\ket{0}+\ket{1} \over \sqrt { 2}}={1 \over 2^{M/2}}\sum_\bi\ket{\bi}_M,
\eeq
where $\bi$ runs over all $2^M$ binary strings of length $M$.
Let $\bt$ denote the binary string $[\QQ \bs,\bs]$.
Let $\P_m^c$ denote the operator that is controlled by the $m$th qubit
$\ket{i_m}$ in the first set $\ket{ \bi }_M$ of $M$ qubits and applies
cNOT operations to the remaining $N-M$ qubits $\ket{t}_{N-M}$ in
positions where the $m$th row of $\PP$ has a 1. Then the final part
of `stage 2' carries out all these operations for $m=1$ to $M$. The effect
of these operators is to add rows of $\tilde \AA =[{\bf I},\PP ]$ to
the binary string $\bt$. Thus the operation of the final stage is:
\beq
 {1 \over 2^{M/2}}\sum_\bi\ket{\bi}_M\ket{\bt}_{N-M}
 \rightarrow
 {1 \over 2^{M/2}} \sum_\bi
   \left[\prod _{m=1}^{M} \P_m^c\right]\ket{\bi}_M\ket{\bt}_{N-M}
 ={1 \over 2^{M/2}} \sum_{{\bf r} \in {\cal C}^\perp(\AA)} \ket{{\bf r}+\bt}_N,
\eeq
and we have generated an encoded state of the form
 (\ref{encodedstate}). If we start from $\sum_{\bs} \alpha_{\bs}
 \ket{\bs}_{\K}$ we get the right
 hand side of \eqref{general}, the most general
 codeword.

\else
\subsection{Decoding and encoding circuits}
 The syndrome of a stabilizer code can be efficiently measured using
 standard quantum gates. General principles for designing
 circuits to encode a quantum state are described in \citeasnoun{NielsenChuang}.
 For CSS codes
 these circuits are particularly straightforward,
 as described in the extended version of the present paper
 \cite{mackaymitchisonmcfadden2003Long}.
\fi

\section{Sparse-graph codes}
\label{djcmsection1}
 It has been proved that there exist  quantum codes with non-zero rate $R$
 and blocklength $N$  that can correct any $t$ errors, with $t \propto N$.
 However, {\em practical\/} codes with these properties have not yet
 been presented. To be practical, a quantum code must satisfy two properties.
 First, the associated classical decoding problem must be practically
 solvable. [To be precise, we want a decoding time polynomial in $N$,
 preferably linear in $N$.]
 And second, the $M_Q$ measurements required to implement the error-correction
 mechanism must be feasible: in our view, an error-correction mechanism
 is much more likely to be feasible if every syndrome measurement involves
 only a small
 subset of size $k \ll N$ of the  qubits,
 rather than a size  $k \propto N$ that is required for a generic
 quantum code.


 We therefore study quantum codes that
 are associated with  {\em sparse graphs}.
 In a sparse-graph code, the  nodes in the graph represent the transmitted
 bits and the constraints they satisfy. Any linear code can be described
 by a graph, but what makes a sparse-graph code
 special is that each constraint  involves only a small number of
 variables in the graph.
 The sparseness  has the immediate advantages that
 (1)
 the quantum syndrome can be measured with sparse
 interactions --
 for example, if the quantum syndrome is found by bringing together
 qubits in pairs, a quantum code
 with a sparse graph requires only of order $N$ interactions,
 rather than $N^2$;
 and (2) there are practical decoding algorithms (in particular,
 the sum-product algorithm) for decoding classical codes
 defined on sparse graphs; indeed, codes based
 on sparse graphs are record-breaking codes for classical channels.

\newcommand{\equalnode}{\raisebox{-1pt}[0in][0in]{\psfig{figure=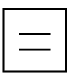,width=8pt}\hspace{0mm}}}
\newcommand{\plusnode}{\raisebox{-1pt}[0in][0in]{\psfig{figure=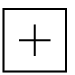,width=8pt}\hspace{0mm}}}
\subsection{Classical sparse-graph codes}
 The archetypal sparse-graph code is
\ifnum \arabic{longversion} > 0 
 \quotecite{Gallager62}
\else
 Gallager's \cite{Gallager62}
\fi
 low-density parity-check code. 
 A {low-density parity-check code}
 is a block code
 that has a parity-check matrix, $\bH$, every row and column of which
 is `sparse'.

\begin{figure}
\begin{center}
\mbox{\raisebox{-0.64in}
{ \bH \hspace{0.02in} =}\hspace{-0.1in}
{\psfig{figure=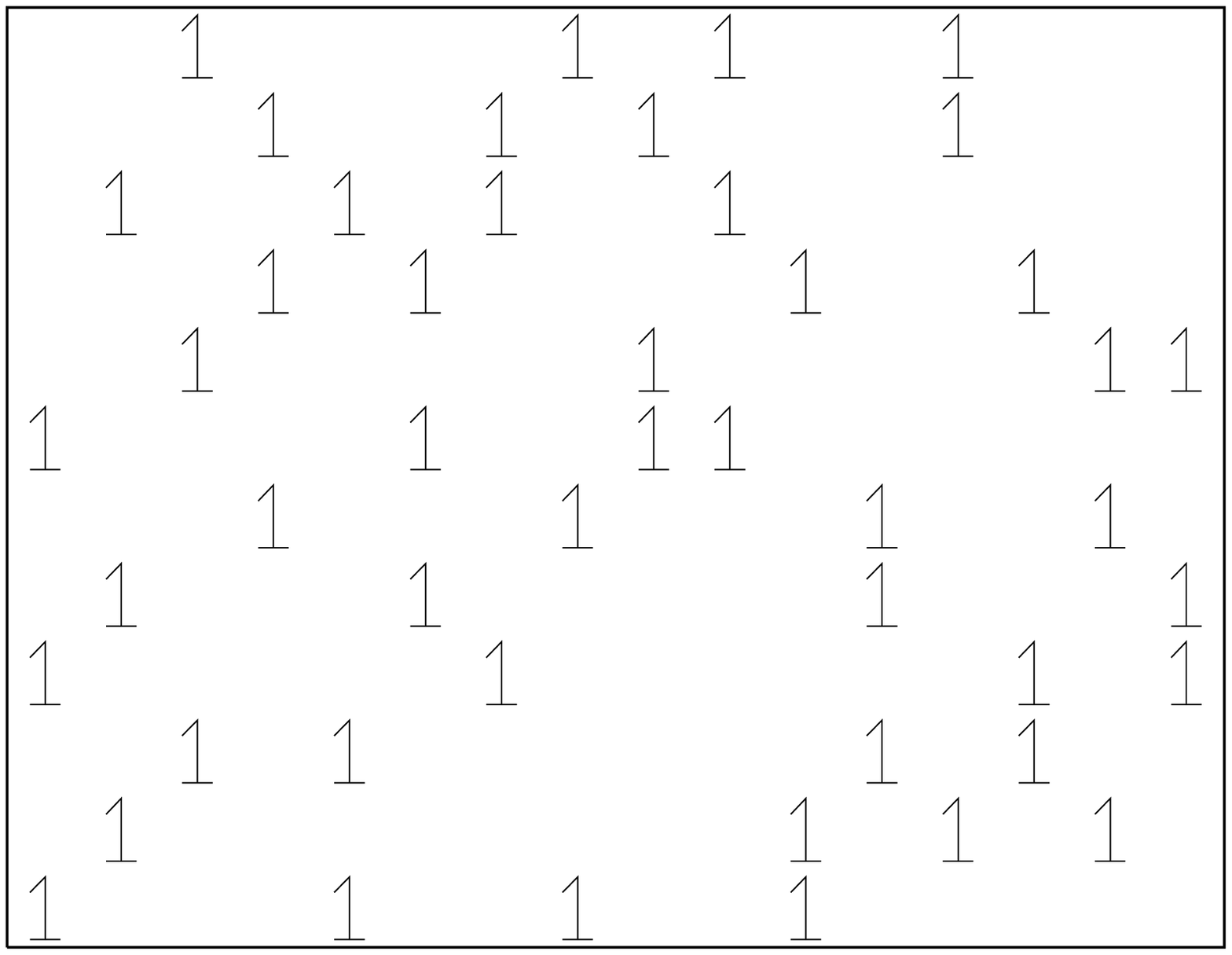,angle=-90,width=1.85in}}
}
 \mbox{
\psfig{figure=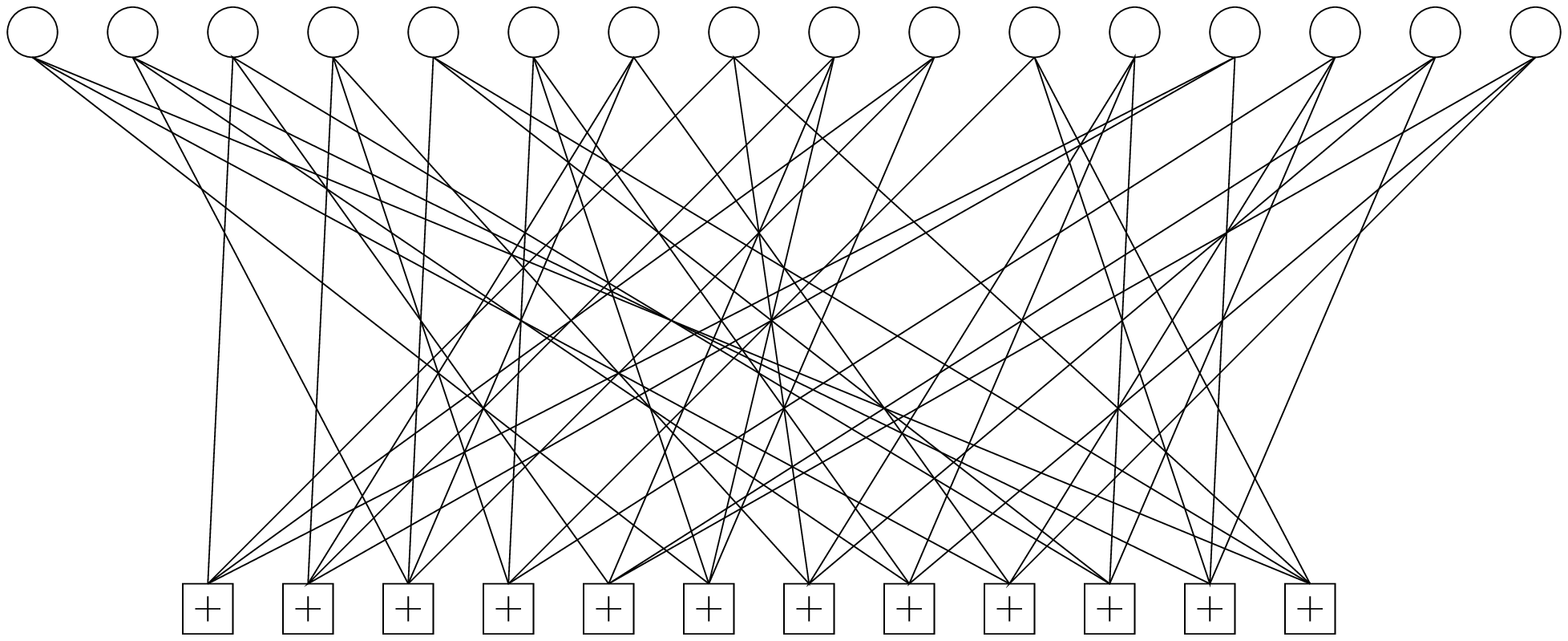,width=1.152in,angle=-90}
}\end{center}
\caption[a]{A low-density parity-check matrix
 and the corresponding graph of  a rate-\dfrac{1}{4}
 low-density parity-check code with
 blocklength $N=16$, and $M=12$ constraints.
 Each white circle represents a transmitted bit, corresponding to
 a column of $\bH$. Each bit
 participates in $j=3$ constraints, represented by
 \plusnode\ squares, corresponding to the 12 rows of $\bH$. Each
 constraint forces the
 sum of the $k=4$ bits to which it is connected to
 be even. From \cite{MacKay:itp}.
}
\label{fig.ldpccb}
\end{figure}

 In a {\em{regular}\/}    low-density parity-check code,
 every column of $\bH$ has the same weight $j$ and every row has the
 same weight $k$; regular   low-density parity-check   codes are constructed at random
 subject to these constraints. 
 A tiny low-density parity-check code with $j=3$ and $k=4$ is illustrated
 in \figref{fig.ldpccb}.
 In spite of their simplicity and sparseness,
 low-density parity-check codes have excellent theoretical and
 practical properties.
 The following results are proved in \citeasnoun{Gallager63} and
 \citeasnoun{mncN}. For any column weight $j \geq 3$,
 low-density parity-check codes
	are  `good' codes, {\em given an optimal decoder\/} (`good'  in the
 technical sense that there exist sequences of codes
 that achieve vanishing error probability at non-zero communication rate).
 Furthermore, they have `good'  minimum
 distance (that is, there exist sequences
 of codes whose 
 distance grows linearly with  blocklength).

 The decoding of a  low-density parity-check code is an
 NP-complete problem \cite{BMT78}, but for practical purposes, as
 long as the graph is sparse, it has been found that
 simple message-passing algorithms can give excellent performance.
 The best such algorithm known is the {sum--product algorithm},
 also known as {iterative probabilistic decoding}
 or {belief propagation}. This iterative algorithm is explained  in
 \cite{mncEL,mncN,MacKay:itp,frey-98}. In one iteration,
 each edge in the graph carries a message (a single real number)
 from its bit node (the circles in \figref{fig.ldpccb})
 to its check node (the squares), representing the relative probability
 of the two states of the bit node; each check node then computes
 the relative
 likelihoods of the two states of each bit node, given the
 incoming messages from the other nodes and the constraint enforced by
 the check node; these  likelihood ratios are sent back up each edge;
 then each bit node collates the likelihoods it receives, multiplying them
 together appropriately in order to compute new probabilities for the next
 iteration. 

 When decoded in this way, regular low-density parity-check codes
 with column weight $j=3$ or $4$
 are quite hard to beat.  Performance even closer to
 the Shannon limit has been achieved by making the sparse graph
 irregular
 \cite{Luby2001b,Richardson98,Richardson2001b}: if the bit nodes have a carefully
 chosen distribution  of degrees, rather than all having the same degree $j$,
 then the effectiveness of the sum-product decoding
 algorithm is enhanced. 
 
 Sparse-graph codes have  great flexibility: we can make
 low-density parity-check codes
 of almost any required blocklength $N$ and rate $R$; and they
 are good codes for a wide variety of channels -- for example
 channels with erasures, channels with a varying noise level and channels with burst noise,
 as well as the traditional binary symmetric channel.

 The challenge that remains is to make a sparse-graph {\em quantum\/} code
 that can correct a good number of errors.
 Our ultimate aim is to find practical codes that
 can correct a number of errors $\propto N$.
 To be precise, we'd like to find a family of codes
 with blocklength $N$
 that can correct {\em almost any\/} $t$ errors,
 where $t \propto N$. The difference between correcting `almost any $t$ errors'
 and `any $t$ errors' is important -- if we wish to approach the Shannon limit,
 we must make systems with the former property.
 In the present paper, we describe first steps
 in this direction.
 The codes we present  do not have distance   $\propto N$;
 nevertheless, they can achieve error-correction at substantial noise-levels,
 and for intermediate values of $N$ (large, but not enormous),
 we believe they are the best known quantum codes.

\subsection{Distance isn't everything}
\label{sec.dist}
 Perhaps we should elaborate on the   perspective on
 code design given above. Much of coding theory, including
 the founding papers of quantum coding theory, emphasizes
 `the number of errors that can be corrected' by a code --
 an emphasis which leads to one trying to make codes that
 have large {minimum distance}.
 However, {\em distance isn't everything}.
 The distance of a code  is of little relevance to
 the question `what is the maximum noise level that
 can be tolerated by the optimal decoder?' Let us explain, assuming
 that the channel in question is
 the standard binary symmetric channel.
 At large blocklength $N$, the Gilbert-Varshamov conjecture (widely believed true)
 asserts that the best binary codes of rate $R$
 have a minimum distance $d$ that satisfies
\beq
	R = 1 - H_2(d/N) .
\eeq
 If we decode such a code with a bounded-distance decoder,
 which corrects `up to $t \simeq d/2$ errors', then the maximum
 noise level that can be tolerated is
\beq
	f^{\max}_{{\rm bdd}} \simeq \frac{d/2}{N} ,
\eeq
 which satisfies
\beq
	R = 1 - H_2( 2 f^{\max}_{{\rm bdd}}  ).
\label{eqbdd}
\eeq
 In contrast, Shannon's noisy channel coding theorem says
 that there exist codes of rate $R$ that when optimally decoded
 give {\em negligible
 probability of error\/}  at noise levels
 up to $f^{\max}_{\rm Shannon}$, which satisfies:
\beq
	R =  1 - H_2( f^{\max}_{\rm Shannon} ).
\label{eqsha}
\eeq
 Comparing (\ref{eqbdd}) and (\ref{eqsha}) we deduce
\beq
	  f^{\max}_{\rm Shannon} = 2  f^{\max}_{{\rm bdd}}  .
\eeq
 The maximum tolerable noise level according to Shannon is twice
 the maximum tolerable noise level for a bounded-distance decoder,
 even if the code has the best  possible distance.
 In order to get close to the Shannon limit, we must tolerate
 a number of errors
 twice as great as  the maximum {\em  guaranteed\/}
 correctable, $t \simeq d/2$.
 Shannon's codes can (asymptotically) correct {almost any} $t_{\rm Shannon}$ errors,
 where $t_{\rm Shannon} =   f^{\max}_{\rm Shannon} N$.

 So, does the minimum distance matter at all?  Well, yes, if our aim
 is to make for a given channel a sequence of codes with {\em vanishing\/} error probability, then
 the minimum distance $d$ must increase along that sequence, since
 the error probability is at least
 $\beta^{\, d},$
 where $\beta$ is a property of the channel independent of blocklength  $N$;
 for example, for the binary symmetric channel, $\beta(f) =  2 f^{1/2} (1-f )^{1/2}$ \cite{MacKay:itp}.
 But the distance does not need to increase {\em linearly} with $N$
 in order for the code to be able to correct {\em almost any\/}
 $t_{\rm Shannon} =   f^{\max}_{\rm Shannon} N$ errors. Moreover,
 if our goal is simply to make a code whose error probability
 is smaller than some figure such as $10^{-6}$ or $10^{-20}$ then there is
 no reason why 
 the distance has to grow at all with blocklength.

\begin{figure}
\footnotesize
\begin{center}
\begin{tabular}{c}
\hspace*{0.2in}\psfig{figure=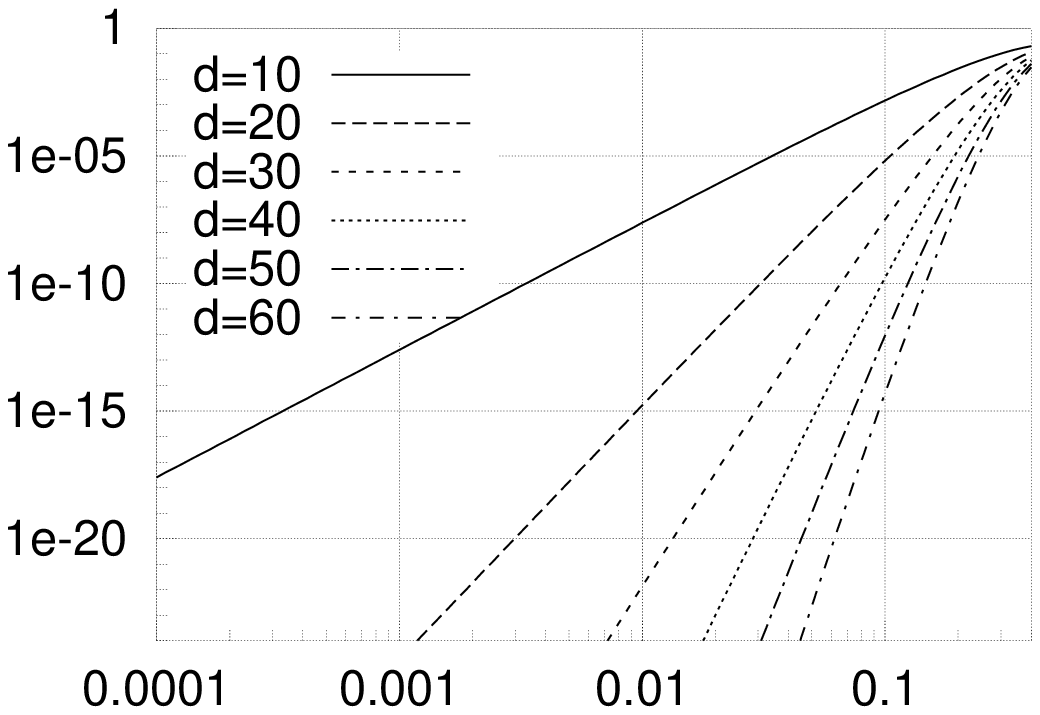,width=1.8in,angle=-90}\\[0.1in]
\end{tabular}
\end{center}
\caption[a]{ The error probability associated with a single codeword of weight $d$, 
${{d}\choose{d/2}} f^{d/2} (1-f)^{d/2}$, as
 a function of $f$.}
\label{fig.dist}
\end{figure}
 This unimportance of the minimum distance for practical purposes
 is illustrated in
 \figref{fig.dist}, which shows
 the error probability associated with a single codeword of weight
 $d$ in the case of a binary symmetric channel with noise
 level $f$.
 From this figure we can see for example that if the raw error probability $f$ is
 about $0.001$, the error probability associated
 with one codeword at distance $d=20$ is smaller than
 $10^{-24}$.

 All else being equal, we might prefer a code with large
 distance, but for practical purposes a code with blocklength
 $N=10,000$
 can have codewords of weight $d=32$ and the error probability
 can remain negligibly small even when the channel
 is creating errors of weight 320.
 Indeed
 we will demonstrate codes with exactly this property.

\section{Dual-containing sparse-graph codes}
\label{djcmsection2}
 In this paper, we focus on   dual-containing codes, which
 have the property that
 every row of the parity-check matrix  is a codeword of the code.

%

\subsection{Some misconceptions}
 We initially thought we would be unable
 to make good quantum codes from dual-containing low-density
 parity-check codes, because
 a  random low-density
 parity-check code is, with high probability, a {\em good\/} code having
 good distance (\ie, distance proportional to $N$),
 whereas the dual of a   low-density
 parity-check code is certainly a {\em bad\/} code having bad distance (its
 distance is at most equal to the row-weight $k$ of the original matrix $\bH$).
 So 
 a low-density
 parity-check code  that contains its dual would have to be a bad code.
 [Bad codes are ones that,  for large $N$,
  cannot  correct a number of errors proportional to $N$.]
 And since almost all  low-density
 parity-check codes are good codes,  low-density
 parity-check codes are not expected to contain their duals.

 This point of view is indeed correct.  The codes that
 we now describe are {\em not\/} good {classical\/} codes;
 they contain  low-weight codewords. However,
 low-weight codewords need not necessarily harm the {\em quantum\/}
 code: if the low-weight codewords
 are  themselves  all contained in the dual of the code,
 then they do not produce quantum errors -- the quantum codewords are
 invariant under addition of codewords contained in the dual.

 Furthermore, from a practical point of view, it is not essential
 to have good minimum distance.  As long as we have a near-optimal
 decoder  (\ie, one that
 works well beyond half the minimum distance of the code), the
 error-probability associated with low-weight codewords
 can be very small, as illustrated by classical turbo codes, which
 have good practical performance even though they typically
 have a few codewords of low weight. A few low-weight codewords  hurt
 performance only a little, because the noise has so many directions
 it could take us in that it is unlikely to take us in
 the few directions that would give rise to confusion with the 
 nearest codewords.

 So while our ideal goal  is to make
 a good  dual-containing code (based on a sparse graph)
 that has no low-weight codewords  not in the dual,
 we are also happy with  a lesser aim: to make
 a  dual-containing code 
 whose low-weight codewords not in the dual are of sufficiently
 high weight that they contribute negligibly
 to the quantum code's error probability.

\subsection{Do dual-containing low-density parity-check codes exist?}

 A code is a regular dual-containing $(j,k)(N,M)$ low-density parity-check code
 if it has an $M\times N$ parity check matrix $\bH$ such that
\ben
\item
 every row has weight $k$ and every column has weight $j$;
\item
 every pair of rows in $\bH$ has an even overlap,
 and every row  has even weight.
\een

 The speculation that motivated this research
 was   that for small fixed $j$ and arbitrarily
 large $N$ and $M$,
 dual-containing $(j,k)(N,M)$ low-density parity-check codes exist
 and almost all of them define good  quantum codes.

\begin{figure}
\[
\framebox{\hspace*{-0.5mm}\raisebox{-0.5mm}{\psfig{figure=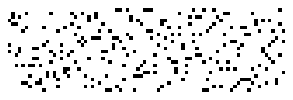,width=3in}}\hspace{0.5mm}}
\]
\caption[a]{A regular dual-containing (3,10)(80,24) low-density parity-check code's parity check matrix,
 found by a Monte Carlo search.}
\label{fig1}
\end{figure}

 Finding such codes proved difficult however.
 \Figref{fig1} shows one code  found by a Monte Carlo search.
 If we increase either the blocklength or the column weight $j$,
 our Monte Carlo searches become
 ineffective.
 [We exclude from our search trivial  solutions
 whose graphs are not well-connected, such as codes
 defined by block-diagonal matrices.]

%

\subsection{A counting argument}
 We count approximately the number of
 regular dual-containing $(j,k)(N,M)$ low-density parity-check codes
 by finding the probability that a randomly created matrix $\bH$ with column-weight
 $j$ satisfies all the $M \choose 2$ self-orthogonality constraints.
 [In defining this ensemble, we neglect the condition that the row weights
 should be exactly $k$, in order to make the calculation easier; we do
 not expect this omission to have any significant effect on the result \cite{LitsynEnumerator}.
 Similar counting arguments for classical {\ldpcc}s give valid predictions.]

 The number of matrices with column-weight $j$ is
\beq
	{\cal N}_{0} = { M \choose j }^N .
\eeq
 As each column of the matrix is created, we can imagine keeping track of
 the ${\cal M} \equiv {M \choose 2}$ inner products (modulo 2) between rows.
 Each column of weight $j$ flips the parity of $j \choose 2$ of the
 inner products. We can view this process as creating a
 random walk on the ${\cal M}$-dimensional hypercube, starting from the origin.
 As each column is added, ${j \choose 2}$ individual steps are taken
 on the hypercube.
 The probability that, after all $N$ columns are created,
 all ${\cal M}$ inner products are zero (the dual-containing condition)
 is the probability that the random walk on the hypercube is at the origin
 after $r = N {j \choose 2}$ steps.
 The probability of return to zero after $r$ steps
 is approximately (for $r \ll {\cal M}$  \cite{mncN})
\beq
	\frac{1}{{\cal M}^{r/2}} \frac{r!}{2^{r/2} (r/2)!}  \simeq  \left(\frac{r/2}{{\cal M}}\right)^{r/2} ,
\label{eq.poo}
\eeq
 so with $r = N {j \choose 2}$ and ${\cal M} = {M \choose 2}$,
\beqan
	P(\mbox{dual-containing}) &\simeq&    \left(\frac{ N {j \choose 2} /2}
                                                             {{M}^2/2 }    \right)^{ N {j \choose 2} /2}
\\ &=&   \left(\frac{ N j(j-1)/2  }
                     {{M}^2 }    \right)^{ N j(j-1)/4 } .
\eeqan
 [\Eqref{eq.poo} can be motivated 
 by imagining that  $r/2$ steps are allowed to select
 distinct dimensions for `outbound' steps,
 and the  other $r/2$  steps are required
 to undo all these steps in a given order
 (which has probability
 $\linefrac{1}{{\cal M}^{r/2}}$);
 the number of ways of ordering these steps is $r!/(2^{r/2} (r/2)!)$.]
 So the number of dual-containing matrices is
\beq
{\cal N}_1=	{\cal N}_0 P(\mbox{dual-containing}) \simeq
	\frac{M^{jN}}{(j!)^N}   \left(\frac{ N j(j-1)/2  }
                                             {{M}^2 }    \right)^{ N j(j-1)/4 } .
\eeq
 What really matters here is whether this quantity scales as $N^{+\alpha N}$
 or as $N^{-\alpha N}$. Are there lots of dual-containing codes,
 or negligibly many?
 We thus drop all terms other than $N$ and $M$, and we assume $N \propto M$.
\beq
{\cal N}_1 \sim 	
	{N^{jN}}   \left(\frac{ 1  }
                              {{N} }    \right)^{ N j(j-1)/4 }
 \sim N^{ N ( j - j(j-1)/4 ) } .
\eeq
 This expression grows with $N$ if
\beq
	 j - j(j-1)/4  > 0 ,
\eeq
\ie\ if $j<5$.

\subsection{Second counting argument}
 We give a second argument for the difficulty of making these dual-containing codes,
 based on a particular construction method.

 A concrete method for making $(j,k)$ low-density parity-check
 codes is to create $j \times k$ permutation matrices
 $\{ \bR_{hi} \}$
 (\ie\ square matrices
 with one 1 per row and one 1 per column)
 of size $(M/j)\times(M/j)$,
 and arrange them
 in the manner illustrated here for the case $(j,k)=(3,4)$:
\beq
	\bH = \left[ \begin{array}{cccc}
		\bR_{11} & \bR_{12} & \bR_{13} & \bR_{14} \\
		\bR_{21} & \bR_{22} & \bR_{23} & \bR_{24} \\
		\bR_{31} & \bR_{32} & \bR_{33} & \bR_{34}
	\end{array}
	\right] .
\label{eq.g.4.3.R}
\eeq
 [While this form of matrix might seem restrictive compared with
 a free choice from all matrices $\bH$
 with column-weight $j$ and row-weight $k$,
 the information content of a matrix  of the form (\ref{eq.g.4.3.R})
 (\ie, the log of the  number of matrices)
 is to leading order equal to the information content of a freely chosen
 matrix with the same $j$ and $k$.]
 We might hope to enforce all the orthogonality rules by picking
 permutation matrices and picking a set of constraints
 which are  for example  of the
 form
\beq
	\bR_{11}  \bR_{12}^{\T}  \bR_{22} \bR_{21}^{\T} = \bI,
\:\:\:\:
	\bR_{11}   \bR_{13}^{\T}  \bR_{33} \bR_{31}^{\T} = \bI,
\:\:\:\:
	\bR_{21}   \bR_{24}^{\T}  \bR_{34} \bR_{31}^{\T} = \bI ;
\label{eq.constraintexample}
\eeq
 these three constraints ensure that all overlaps in the first group of columns are
 compensated by  overlaps elsewhere in the matrix.
 The total number of such constraints required to enforce
 the dual-containing property would be $k {{j}\choose{2}} /2$.
 The number of degrees of freedom in defining a matrix like (\ref{eq.g.4.3.R})
 (where one
 degree of freedom is the freedom to choose one permutation matrix)
 is $(k-1)(j-1)$, since without loss of generality, the top row and
 left column can all be set to $\bI$.
 So a construction like this only has freedom to make a variety of
 codes if
\beq
	(k-1)(j-1) > k {{j}\choose{2}} /2 ,
\eeq
\ie,
\beq
	(k-1) > k {j} / 4 
\eeq
 which cannot be satisfied if $j \geq 4$,
 and can be satisfied if $j=3$ and $k>4$.

 These two results are bad news for our mission. We had hoped to
 be able to make numerous random codes with $j=5$. We viewed
 $j \geq 5$ as a necessary constraint in order that the
 graph of the code have good expansion properties.
 A dual-containing code with $j=4$ would, we believe, be bound to
 have low-weight codewords (rather as regular Gallager codes
 with $j=2$ have low-weight codewords).

 Nevertheless, we have found a few ways to
 make small numbers of dual-containing codes, which we now describe.
 Furthermore, the negative arguments
 are only bad news for {\em dual-containing\/} sparse-graph codes;
 they do not rule out the possibility  of making more general  sparse-graph codes
 that satisfy the  twisted product
 constraint (\ref{twistzero}).


\subsection{Making dual-containing low-density parity-check codes}
 We have tried three approaches to constructing  codes satisfying
 the even-overlap constraint:
 random constructions using Monte Carlo search;
 constructions in which the even-overlap constraints are
 deliberately built into the {\em local\/} structure of the sparse graph;
 and constructions in which the constraints are satisfied by a
 deliberate choice of {\em global\/} structure. Only  the
 third of these approaches worked out for us.
 The codes we make are slightly irregular in that while the row weights
 are all equal to $k$, the column weights are slightly non-uniform.

\subsection{List of constructions of dual-containing sparse-graph codes}
\label{subsec.main}
\begin{figure}\small
\begin{center}
\raisebox{0in}[0in]{\mbox{\psfig{figure=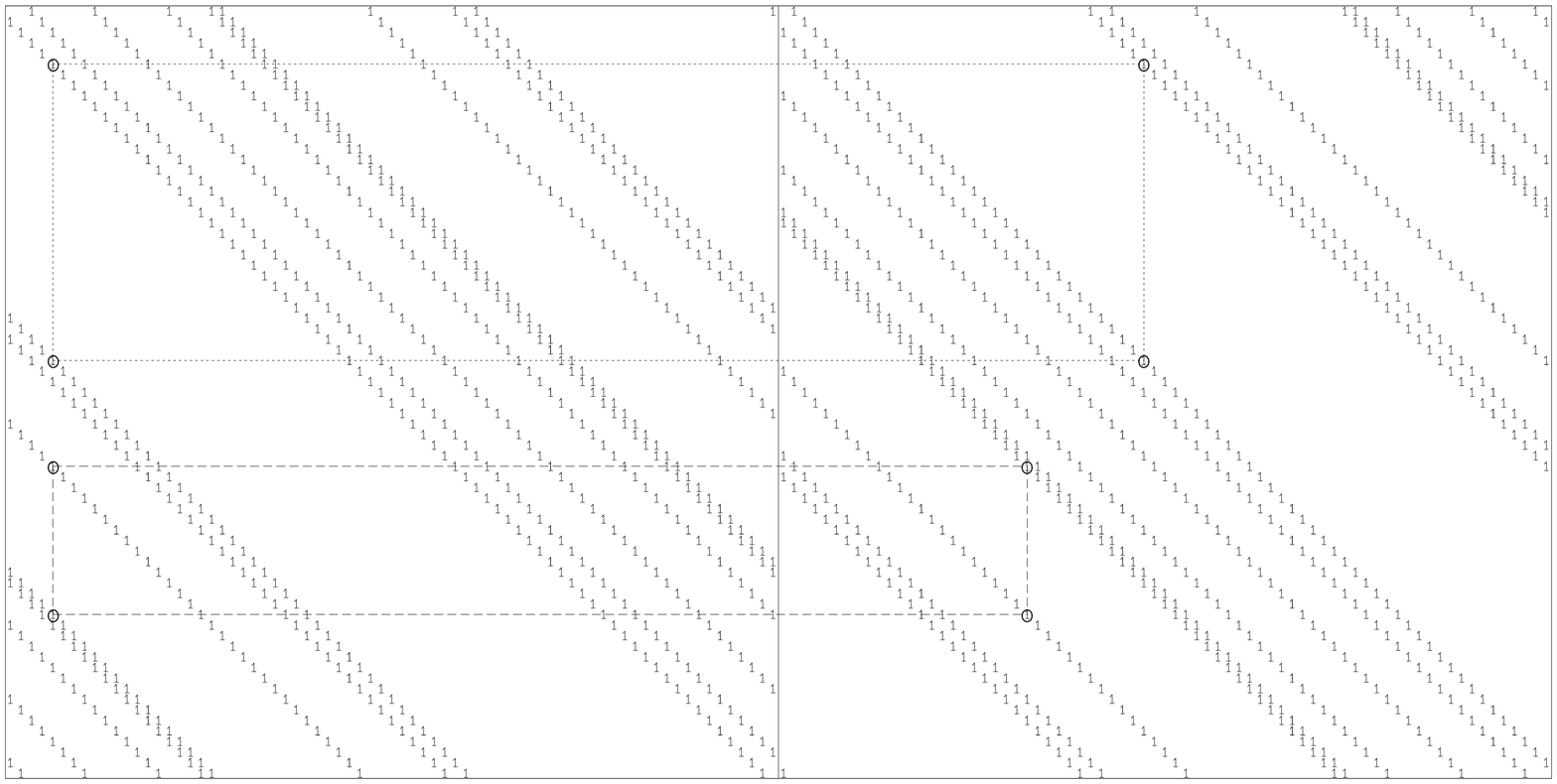,width=3in,angle=-90}}}
\end{center}
\caption[a]{
 Example of a  matrix $\bH_0  = \left[ \matrix{ \bC ,  \bC^{\T} } \right]$
 used in construction B.
 The dotted lines illustrate the way in which two vertical differences
 in column 5 of the left hand matrix
 are reproduced in the right half of the matrix.

 To make a  code with classical rate $3/4$ and
 quantum rate $1/2$, we can take the top half of this matrix.
}
\label{fig.conB}
\end{figure}

 In this paper we present four
 constructions of sparse-graph codes,
 all based on sparse cyclic matrices. We call the constructions B, U, N, and M.
 Of these, the first is the most successful.
\ben
\item
{\bf  Construction B}: `B' is mnemonic for bicycle.
 To make a {\bf Bicycle code} with row-weight $k$, blocklength $N$, and number of
 constraint nodes $M$, 
 we take a random sparse $\mdfrac{N}{2} \times \mdfrac{N}{2}$ cyclic matrix $\bC$ with row-weight $k/2$,
 and define 
\beq
	\bH_0 = \left[ \matrix{ \bC ,  \bC^{\T} } \right] . 
\eeq
 By construction, every pair that appears in $\bC$ appears
 in $\bC^{\T}$ also (\figref{fig.conB}). 
 Then we delete some rows from $\bH_0$ to obtain
 a matrix $\bH$ with $M$ rows.
We delete rows using the heuristic
 that the column weights of $\bH$ should be as uniform as possible.
We usually choose the non-zero entries in $\bC$
 using a difference set satisfying the property that every difference (modulo $N/2$)
 occurs at most once in the set.  

 This construction has the advantage that the choice of $N$, $M$, and $k$
 is completely flexible.  


 The disadvantage of this construction is that the deleted rows are all low-weight codewords
 of weight $k$, which  are unlikely to be in the dual.
 Thus technically speaking we cannot make `good' codes in this way --
 if we fix $k$ and increase the blocklength, the error probability will not vanish (\cf\ section \ref{sec.dist});
 nevertheless for useful blocklengths, we will show that the
 error probability associated with these low-weight codewords
 is negligible, for practical purposes.


\begin{boxfloat}
\begin{framedalgorithmw}{\ourboxwidth}
 An example of a  perfect difference set is
 the  set of integers
\[
 0,   3,   5,  12 \:\:  ({\rm mod}\,\,13).
\]
 A perfect difference set on the additive group of size $M$
 has the property that
 every integer from 0 to $M-1$ can be written as the difference
 of two integers in the set (modulo $M$) in exactly one way.

 For example, the differences between the above integers, modulo 13, are
\[
\begin{array}{lccclccclccclcc}
    & &{\bf  }&&
 3-0&=&{\bf 3}&&
 5-0&=&{\bf 5}&&
12-0&=&{\bf 12}\\
0-3&=&{\bf 10}&&
&&&&
5-3&=&{\bf 2}&&
12-3&=&{\bf 9}\\
0-5&=&{\bf 8}&&
3-5&=&{\bf 11}&&
   &&{\bf }&&
12-5&=&{\bf 7 }\\
0-12&=&{\bf 1}&&
3-12&=&{\bf 4}&&
5-12&=&{\bf 6}&&
\end{array}
\]

 Some of our constructions use plain {\sl difference sets\/}
 in which every difference occurs either one or zero times.

\end{framedalgorithmw}
\caption[a]{%
 Perfect difference sets.
 An entertaining tutorial on difference sets
 by Kris Coolsaet can be found at {\tt http://www.inference.phy.cam.ac.uk/cds/}.
}
\label{boxCDS}
\end{boxfloat}

\begin{figure}\small
\begin{center}
\begin{tabular}{cc}
\begin{tabular}{c}
\mbox{\psfig{figure=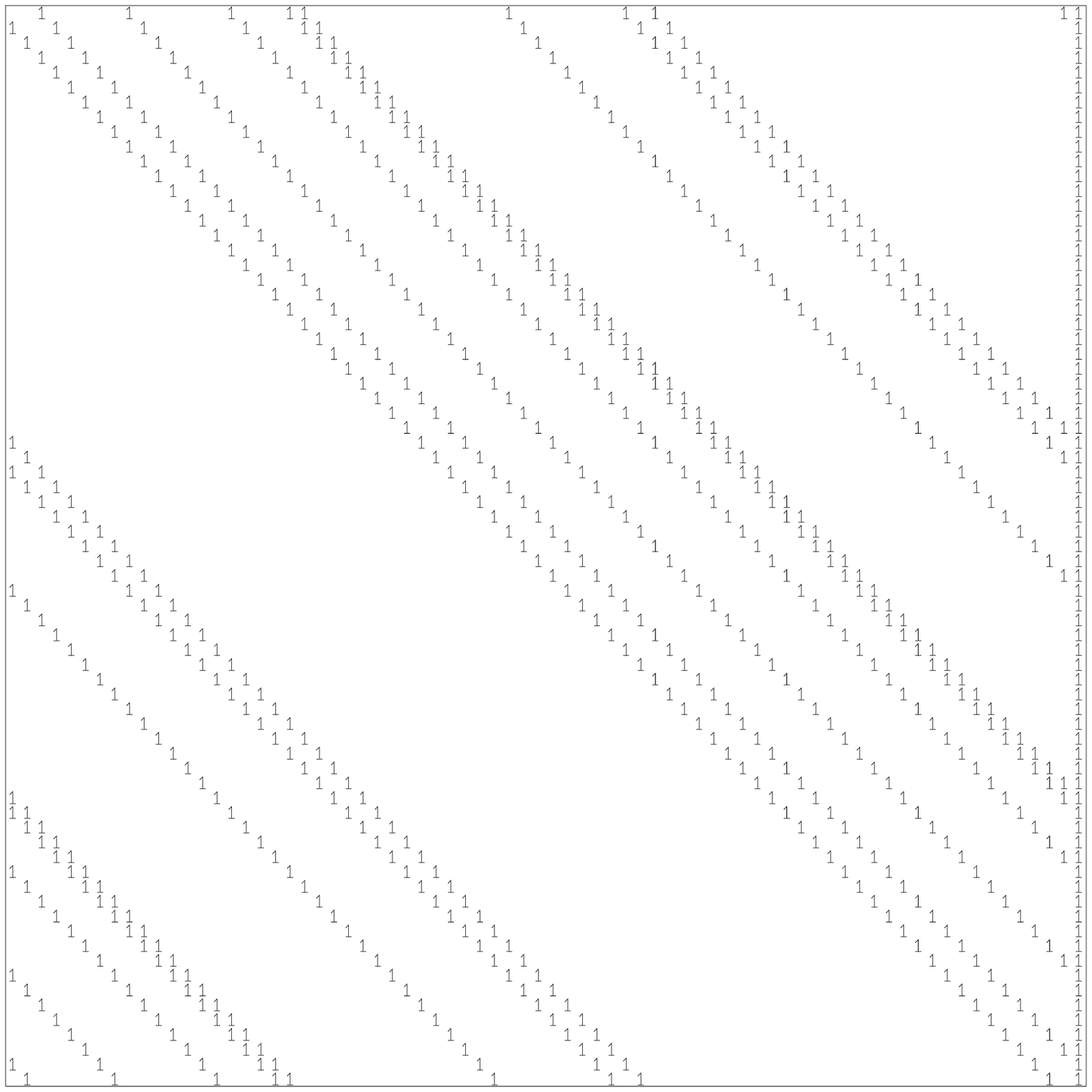,width=1.5in,angle=-90}}
\\
\end{tabular}
&
\begin{tabular}{c*{5}{r}}
\multicolumn{6}{c}{\sc difference-set cyclic codes} \\ \hline
$N$		& 21	& 73	& 273	& 1057	& 4161\\
$M$		&{\bf 10}&{\bf 28}&{\bf 82}&{\bf 244}&{\bf 730}\\
$K$		& 11	& 45	& 191	& 813	& 3431\\
$d$		&{\bf 6}& {\bf 10}&{\bf 18}& 34	& 66\\
$k$		& {\bf 5}& {\bf 9}& 17	& 33	& 65\\
\end{tabular}
\\
(a)& (b) \\
\end{tabular}
\end{center}
\caption[a]{(a) The parity check matrix of the `Unicycle' code of blocklength $N=74$.
 (b) Table of  five difference-set cyclic codes from which
 useful unicycle codes may be derived.}
\label{fig.dsc}
\end{figure}

\item
{\bf Construction U}: `U' is mnemonic for unicycle.
{\bf Unicycle codes} are made with the help of a perfect difference
 set (\boxref{boxCDS}) over the additive group of size $\tilde M=73$, $273$, $1057$, or $4161$.
 For example, a perfect difference set for the group of size $\tilde M=73$
 is $\{ 2,8,15,19,20,34,42,44,72\}$.
 $\tilde M$ will be the number of rows in the code's parity-check matrix.
 By making a cyclic matrix $\bC$ from a perfect  difference set,
 we obtain a parity check matrix that defines a code with
 blocklength $N=\tilde M$ and 
 a number of independent parity constraints $M$ given in the
 table of \figref{fig.dsc}(b), which
 also shows the distance $d$ of each code and the row-weight $k$.
 The  number of independent parity constraints is much smaller than the total
 number of rows $\tilde M$.

 Difference set cyclic codes have been found to have impressive
 performance on classical channels when decoded by
 message passing decoders \cite{Karplus1991,Lucas99}.

 The perfect-difference-set property implies that
 all pairs of rows of $\bC$ have an overlap of 1.
 To create a dual-containing code, we need to make these overlaps
 even. We do this by extending the parity check matrix, adding
 one extra column, all ones. Since all pairs of
 rows of this additional column have an overlap of 1,
 all rows of the new parity check matrix have overlap 2.
 Thus we have defined a dual-containing ($N+1,K+1$)
 code with
 parity check matrix $\bH$ whose row weight is $k+1$ (for example,
 10, for $\tilde M=73$) and whose column weights are all $k$
 except for one column with enormous weight $\tilde M$.
 When decoding, we can handle this one column in a special way.
 We can view the code as  the union of two codes,
 one for each setting of the extra bit.  The first code
 (in which the extra bit is set to zero)
 is the original difference set cyclic code; the second is the code obtained by
 adding the vector $(1\,1\,1\,1\ldots1)$ to all words of the first.
 We can decode each of the two codes separately by the sum-product
 algorithm then (if both
 decoders  return a codeword) select the
 codeword that has maximum likelihood.

 A disadvantage of these codes is that so few of them exist, so
 there is little choice of the parameters $N$, $M$, and $k$.
 Another disappointing property is that the code has codewords of
 weight equal to the row-weight.

\item

 {\bf Construction N} also makes use of cyclic difference sets.  We choose
 a number of rows $M$ and an integer $v$ such as $v=4$ and
 create $v$ cyclic matrices from $v$ cyclic difference
 sets over $\{ 0,1,2,\ldots (M-1)\}$, with the special property
 that every difference (modulo $M$) occurs zero times or twice (preferably once
 in one set and once in another).
 Such sets of difference sets are not easy to find but can be found
 by computer search. An example of $v=4$ sets for $M=500$ is:
\[
\begin{array}{@{\{}*{5}{c}@{\}}}
  0 & 190 & 203 & 345 & 487 \\
 0 & 189 & 235 & 424 & 462 \\
 0 & 94 & 140 & 170 & 310 \\
 0 & 15 & 47 & 453 & 485 \\
\end{array}
\]
 with an illustrative pair of matched differences
 being $170-140 = 30$ (in set 3) matching $15-485 =30$ $({\rm{mod}}\,500)$ (in set 4).

 We turn each set into a square cyclic matrix and put
 the matrices alongside each other to make the parity check matrix $\bH$.
 Since every difference occurs an even number of times, the overlap between
 any pair of rows is even.


\item

 {\bf Construction M} is a special case of construction N  in which
 all the $v$ difference sets are derived from a single parent
 different set with $w$ elements. The parent difference set has the property that
 every difference (modulo $M$) occurs one or zero times.
 We  select the $v$ difference sets  using a design (for example, for $v=8$ and $w=14$,
 a $14,7$ quasi-symmetric design) that ensures every pair appears
 in exactly two derived difference sets. [We thank Mike Postol for
 providing this quasi-symmetric design.]
 The $v$ derived difference sets define $v$ square cyclic matrices
 each of which is either  transposed
 or not when they are put alongside each other to define $\bH$.

 For the codes reported here we used the  $14,7$ quasi-symmetric design
 shown in \tabref{qsd}.
\begin{table} \footnotesize
\[\footnotesize
\begin{array}{c*{7}{p{0.23in}}c}
\mbox{Matrix number} & \multicolumn{7}{c}{\mbox{Elements selected}} & \\ \midrule
1:&	4  &     6  &     7 &      8   &    9   &    10  &    12&   \\
2:&	1  &     5  &     7 &      9   &    10  &    11  &    13& \T\\
3:&	1  &     2  &     6 &      10  &    11  &    12  &    0 &   \\
4:&	2  &     3  &     7 &      8   &    11  &    12  &    13& \T\\
5:&	1  &     3  &     4 &      9   &    12  &    13  &    0 &   \\
6:&	2  &     4  &     5 &      8   &    10  &    13  &    0 & \T\\
7:&	3  &     5  &     6 &      8   &    9   &    11  &    0 &   \\
8:&	1  &     2  &     3 &      4   &    5   &    6   &    7 & \T\\
\end{array}
\]
\caption[a]{The $14,7$ quasi-symmetric design.
 The final column
 indicates by $\T$ the matrices that were transposed.
}\label{qsd}
\end{table}
 We used parent difference sets over $M=273$ and $M=1901$.

 Constructions $N$ and $M$ both suffer from two problems.
 First, a code built from $v$ square cyclic matrices each of weight $j$
 inevitably has many  codewords of weight $2j$: for any pair of square cyclic matrices
 associated with vectors ${\bf f}$ and $\bg$ (both of weight $j$), the vector $(\bg,{\bf f})$ is a
 weight-$2j$ codeword,
 as is any  vector built from cyclic shifts of $\bg$ and $\bff$.
 Second, every square cyclic matrix has `near-codewords' associated with it.
 We define a {\em $(w,v)$ near-codeword\/} of a code with parity-check 
 matrix $\bH$ to be a vector $\bx$ with weight $w$
 whose syndrome $\bz(\bx) \equiv \bH \bx$ has weight $v$.
 Near-codewords with both small $v$ and relatively small $w$
 tend to be error states from which the sum-product decoding algorithm
 cannot escape. A small value of $w$ corresponds
 to a quite-probable error pattern, while the small value of
 $v$  indicates that only a few
 check-sums are affected by these error patterns.
 \cite{MacKayHighRate98,MacKayPostol2002}.
 If the square matrix is defined by a vector $\bg$ of weight $j$, then the
 component-reversed vector $\tilde \bg$ is a $(j,j)$ near-codeword.


\een

\section{Channel models chosen for simulations}
\label{djcmsection3}
 When comparing candidate codes for quantum error-correction,
 we have studied the codes' performance on three classical channels.
 Because of the correspondence between classical codes and stabilizer
 codes (section \ref{sec2.2}), our results will give bounds on the performance
 of the quantum codes
 defined in the previous section. To each classical channel there
 corresponds a quantum channel whose error operators $\{E_\alpha\}$ are
 related to the binary noise vectors $\{e_\alpha\}$ of the classical
 channel as described in section \ref{sec2.2}.  

 {\bf Two binary symmetric channels}.
 The simplest channel models $\X$ errors and $\Z$ errors
 as independent events, identically distributed with a flip probability
 $\fm$. The suffix `m' denotes the `marginal' flip probability.
 Most of our simulations have used this model because
 it allows easy comparison with textbook codes.

 However, a possibly more realistic channel is the
\ifnum \arabic{longversion} > 0 
 {\bf depolarizing
  channel} (\eqref{eq.depol}).
 The classical analogue that is
 simulated here is the 
 {\bf 4-ary symmetric
 channel},  which creates $\X$ errors, $\Y$ errors,
 and $\Z$ errors with equal probability $f/3$.
\else
 {\bf depolarizing
  channel}, which  creates $\X$ errors, $\Y$ errors, and $\Z$
 errors with equal probability $f/3$.
 The classical analogue that is
 simulated here is the 
 {\bf 4-ary symmetric
 channel}.
\fi
 The total
 flip probability is $f$. The probability of no error is $1-f$.
 If we describe a $\Y$ error
 as the combination of an $\X$ error and a $\Z$ error
 then the marginal flip probability (the probability of an $\X$ error,
 ignoring what's happening to $\Z$)
 is
\beq
 \fm = 2 f/3.
\eeq
 The  4-ary symmetric  channel may be treated by a decoder as if  it
 were a pair of binary symmetric channels, but this approximation
 throws away information about the correlations between $\X$ errors and $\Y$ errors.
 The sum-product decoding algorithm can retain this information
 with only a small increase in complexity.

 {\bf Diversity of qubit reliabilities}.\label{sec.logit}
 We also studied a third channel,  with  noise level varying from qubit to qubit,
 to illustrate the gains possible when the decoder can make use
 of  known variations in noise level. We chose the standard Gaussian
 channel, the workhorse  of communication theory,
 to define our classical channel.
 One way to describe this channel is as a binary symmetric channel
 where  the noise level
 for each bit is set by drawing a random variable $y$ from a Gaussian distribution with mean $1$
 and standard deviation $\sigma$, and setting the flip probability to
 $f_n = 1/(1+\exp(2|y|/\sigma^2))$.
 If this  channel is treated by the decoder as  a binary symmetric channel
 then its 
 marginal flip probability is $\fm = \erf( 1/\sigma)$,
 where $\erf (z) = \int_{z}^{\infty} e^{-z^2/2} dz/\sqrt{2\pi}$; but
 neglecting the reliabilities $\{ f_n \}$ in this way reduces the maximum achievable communication rate.

\subsection{Benchmark communication rates for symmetric channels}
 We define three benchmark rates for the classical channels, and
 obtain  benchmark quantum rates from them.

 The capacity of the binary symmetric channel is
\beq
	C_{\rm BSC}(\fm) = 1 - H_2(\fm) .
\eeq
 The capacity, also known as the Shannon limit, is the
 maximum rate at which reliable communication can be achieved over
 the binary symmetric channel with flip probability $\fm$.

 The classical Gilbert rate is defined by
\beq
	R_{\rm GV}(\fm) =  1 - H_2(2 \fm) \:\:\:\: \fm \in (0,\dfrac{1}{4}).
\eeq
 This is widely believed to be
 the maximum rate at which a {\em bounded-distance decoder\/}
 can communicate over the channel.

 The capacity of the classical 4-ary symmetric channel is
\beq
	C_{4}(f) = 2 - ( H_2(f) + f \log_2 3 ) .
\eeq
 For comparability, we rescale this function
 as follows:
\beq
	C_{4B}(\fm) \equiv \half  C_{4}( 3\fm/2 ) .
\eeq
 This is the maximum rate at which reliable  communication can be achieved over
 each half of the $4$-ary symmetric channel whose marginal flip probability
 is $\fm$.

 For each classical rate $R$ we can define a quantum rate $R_Q (R) = 2 R -1$
 (because a classical dual-containing code with $M$ constraints
 and rate $(N-M)/N$ defines a quantum code with rate $(N-2M)/N = 2R-1$).

 Thus we define
\beq
	C^Q_{\rm BSC}(\fm) = 1 - 2 H_2(\fm) ;
\label{eq.CQBSC}
\eeq
\beq
	R^Q_{\rm GV}(\fm) =  1 - 2 H_2(2 \fm) ;
\eeq
 and
\beq
	C_{4}^{Q}(\fm) =  1  - ( H_2(3\fm/2) + (3\fm/2) \log_2 3 ) .
\eeq
 $C^Q_{\rm BSC}$, defined in \eqref{eq.CQBSC}, should not be confused with the quantum
 channel capacity. It is the maximum rate attainable by stabilizer
 codes constructed from dual-containing codes, and therefore provides
 an upper bound (and ideal goal) for this particular class of
 code. $R^Q_{\rm GV}$ and $C^Q_4$ have  similar meanings.
 We also define the quantum GV bound for
 stabilizer codes; this is labelled the stabilizer rate in
 figures that follow:
\beq
	R_{GV 4}^{Q}(\fm) =    1  - (  H_2(2\fm) + (2\fm) \log_2 3 ) , \:\:\: \fm \in (0,\dfrac{1}{6}) .
\eeq

%
\section{Results}
\label{results}
 We measured  the block  error probability  of each code
 as a function of noise level by empirical experiments involving
 many simulated decodings.
 We  simulated the  classical channel
 then attempted to solve the resulting decoding problem using the
 standard decoding algorithm for low-density parity-check codes,
 the sum-product algorithm \cite{Gallager63,mncEL}.
 The outcome of each decoding is either the correct decoding,
 or a block error. Each block error is classified as a
 detected error if the decoder itself identifies
 the block as one that is known to be in error,  and otherwise  as an undetected error
 (if the decoder finds  a valid decoding that is not the correct decoding).
 In the figures we graph the performance of a code by showing
 its total block error probability, with two-sigma error bars.
 The caption of each figure notes  whether the block errors
 were entirely detected errors. 

\subsection{Construction B}
\begin{figure}
\begin{center}
\ebno{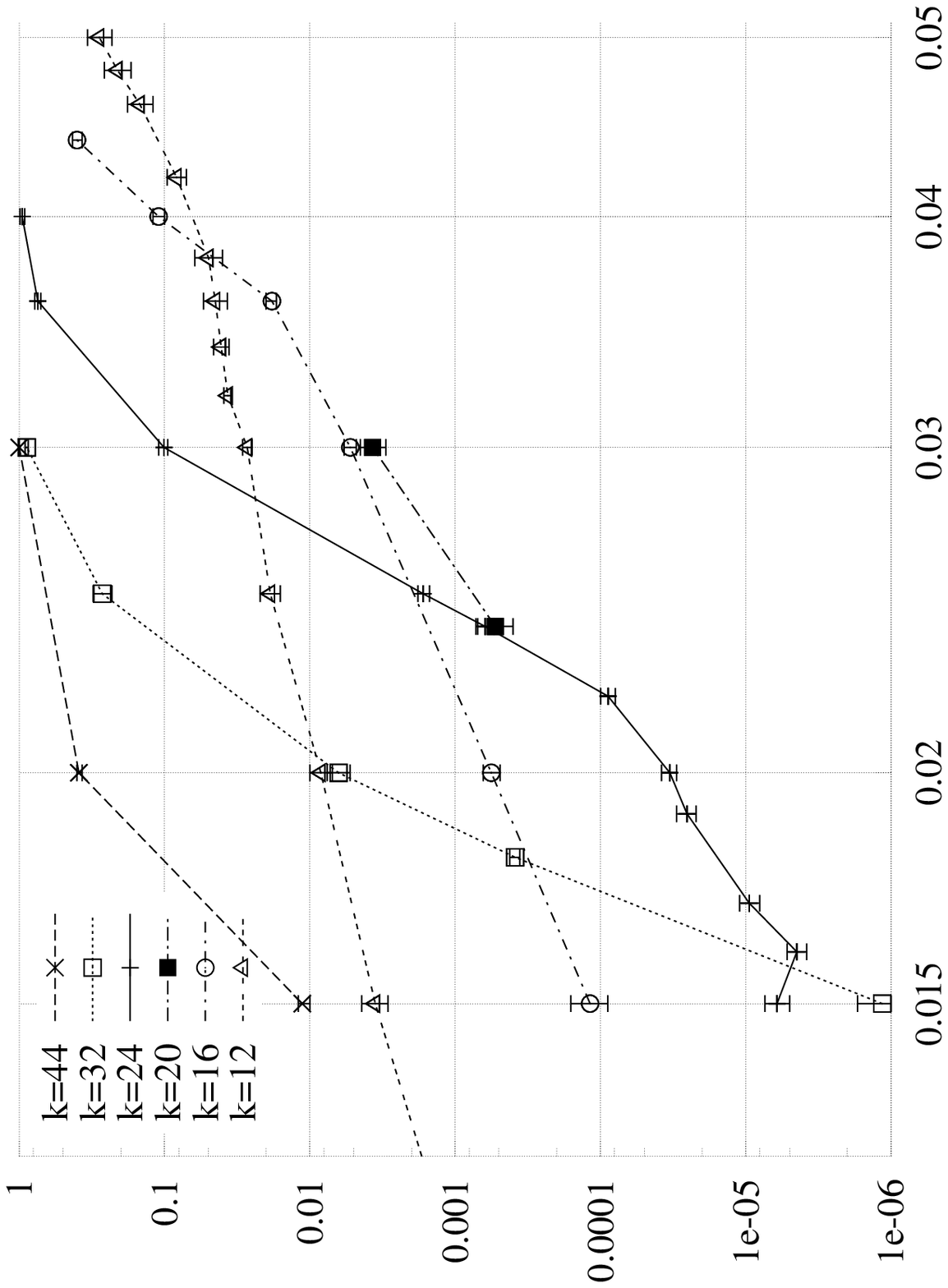}
\end{center}
\caption[a]{Performance of  dual-containing binary codes of construction B with
 parameters $N=3786$, $M=1420$, and row weights ranging from
 $k=88$ to $k=12$, on the binary symmetric channel,
 as a function of the flip probability $\fm$. The vertical axis shows
 the block error probability. All errors were detected errors, that is,
 the decoder reported the fact that it had failed.
 The quantum codes obtained from these codes  have quantum rate $R_Q = 1/4$.
}
\label{fig.3786.1420}
\end{figure}

\begin{figure}
\begin{center}
\ebno{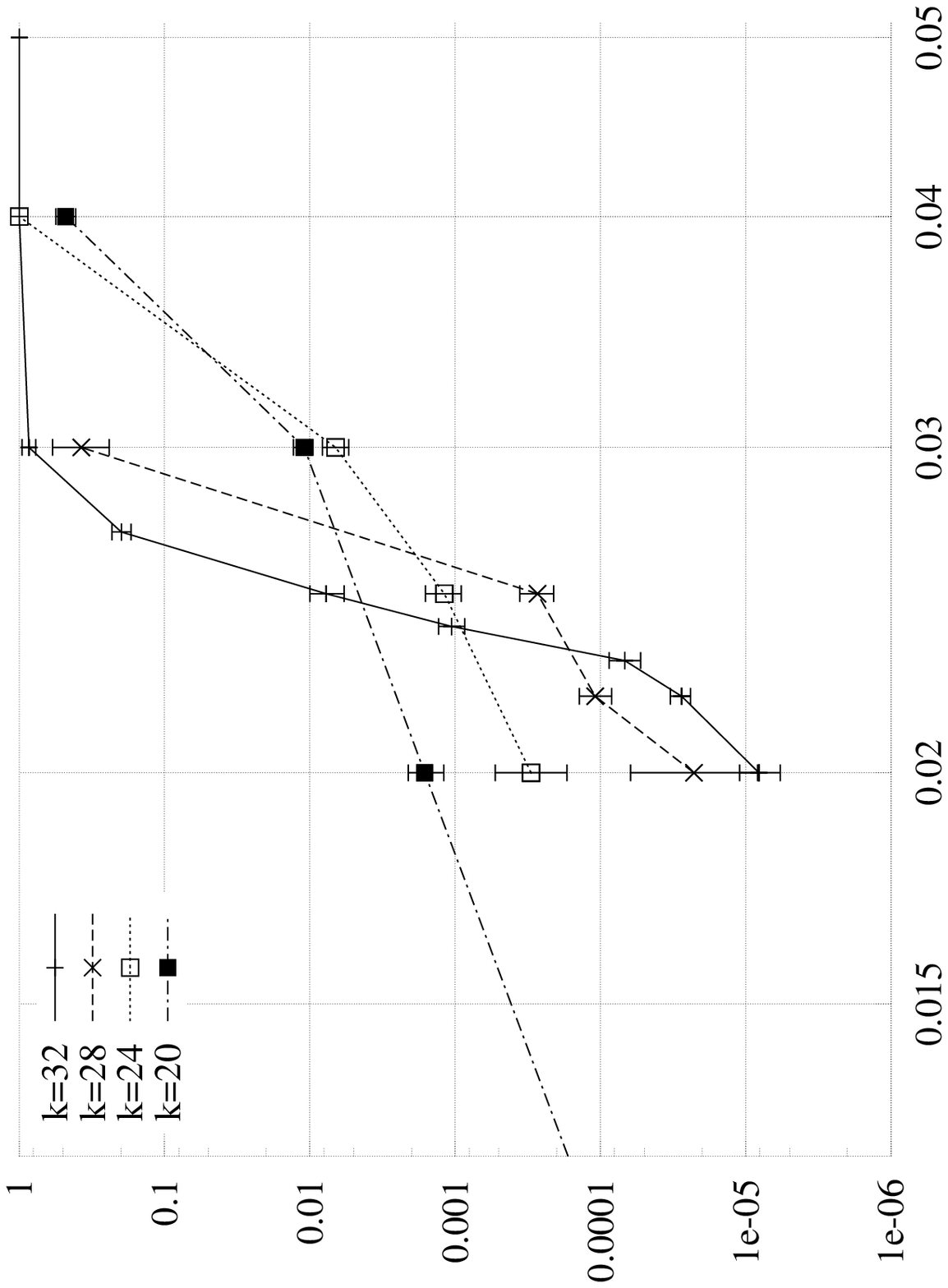}
\end{center}
\caption[a]{Performance of  dual-containing binary codes of construction B with
 parameters $N=19,014$, $M=7131$, and row weights ranging from
 $k=32$ to $k=20$, on the binary symmetric channel,
 as a function of the flip probability $\fm$. The vertical axis shows
 the block error probability. All errors were detected errors, that is,
 the decoder reported the fact that it had failed.
 The quantum codes obtained from these codes  have quantum rate $R_Q = 1/4$.
}
\label{fig.19014}
\end{figure}

\begin{figure}
\begin{center}
\ebno{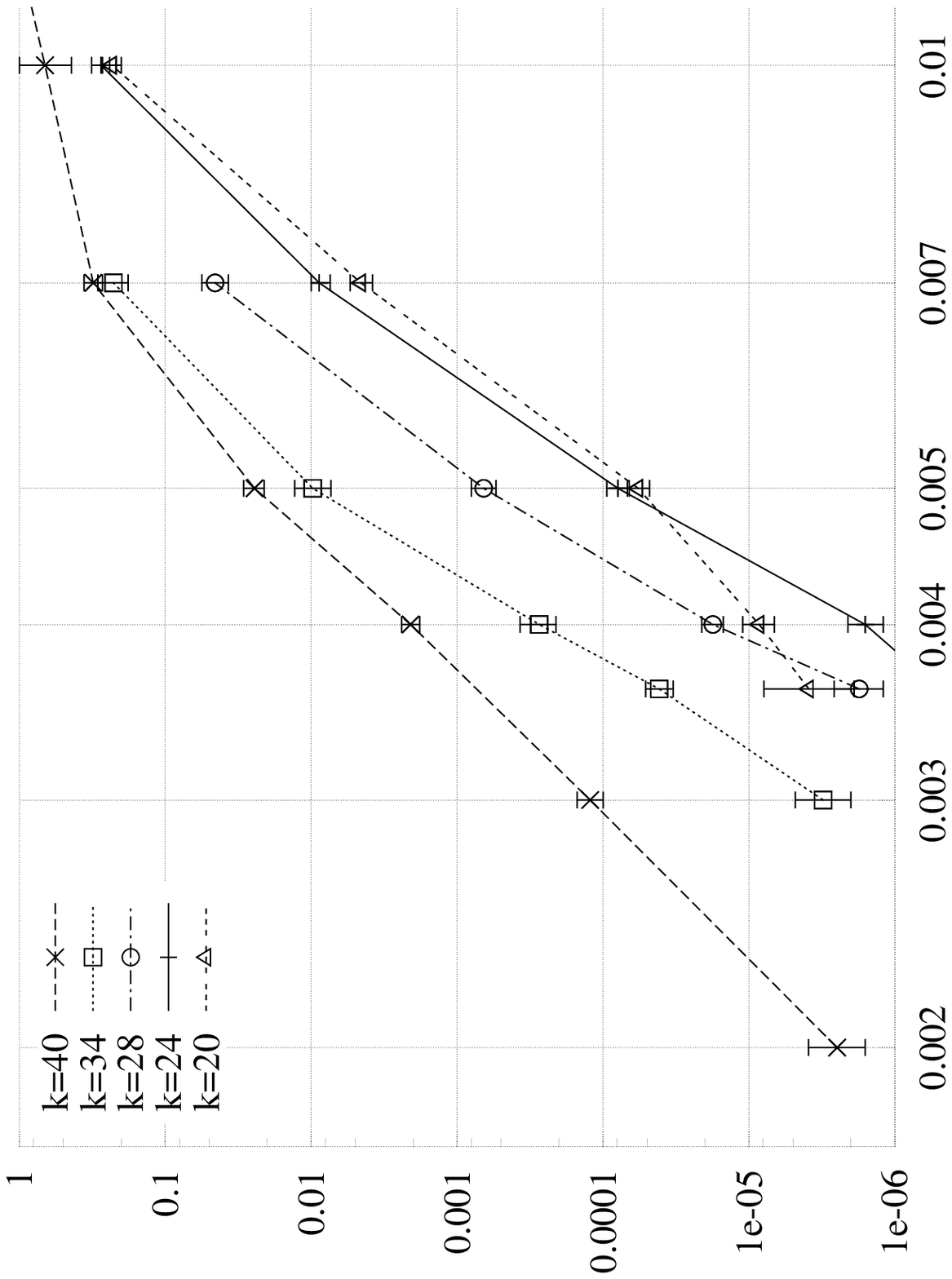}
\end{center}
\caption[a]{Performance of  dual-containing binary codes of construction B with
 parameters $N=3786$, $M=473$, and row weights ranging from
 $k=40$ to $k=20$, on the binary symmetric channel,
 as a function of the flip probability $\fm$. The vertical axis shows
 the block error probability. All errors were detected errors, that is,
 the decoder reported the fact that it had failed.
 The quantum codes obtained from these codes  have quantum rate $R_Q = 3/4$.
}
\label{fig.3786.473}
\end{figure}

 Figures \ref{fig.3786.1420}--\ref{fig.3786.473}
 show results for some codes of construction B.
 In each figure the blocklength and rate are fixed,  and
 the different codes have different row-weights $k$.

 By comparing  \figref{fig.3786.1420}
 and \figref{fig.19014} one may see the effect of increasing the
 blocklength from $N=3786$ to $N=19,014$ while
 keeping the quantum rate fixed at $R_Q=1/4$.
 Notice that while the minimum distance of the code with
 $N=3786$ and $k=24$ is known to be at most 24,
 this code is able to correct almost any 80 errors
 with a block error probability smaller than $10^{-4}$;
 similarly the code with $N=19,014$ and $k=32$
 has minimum distance at most 32 yet it
 can correct almost any 380 errors with   a block error probability
 smaller than $10^{-5}$.

 \Figref{fig.3786.473} shows codes with blocklength $N=3786$
 and a larger rate of $R_Q=3/4$.

 For each blocklength and rate there is an optimum row-weight $k$
 for the codes of construction B. For a blocklength of   $N=3786$
 the optimum was about $k=24$; for the larger blocklength
 of 19,014, the optimum
 was about $k=32$.


\begin{figure}
\begin{center}
\ebno{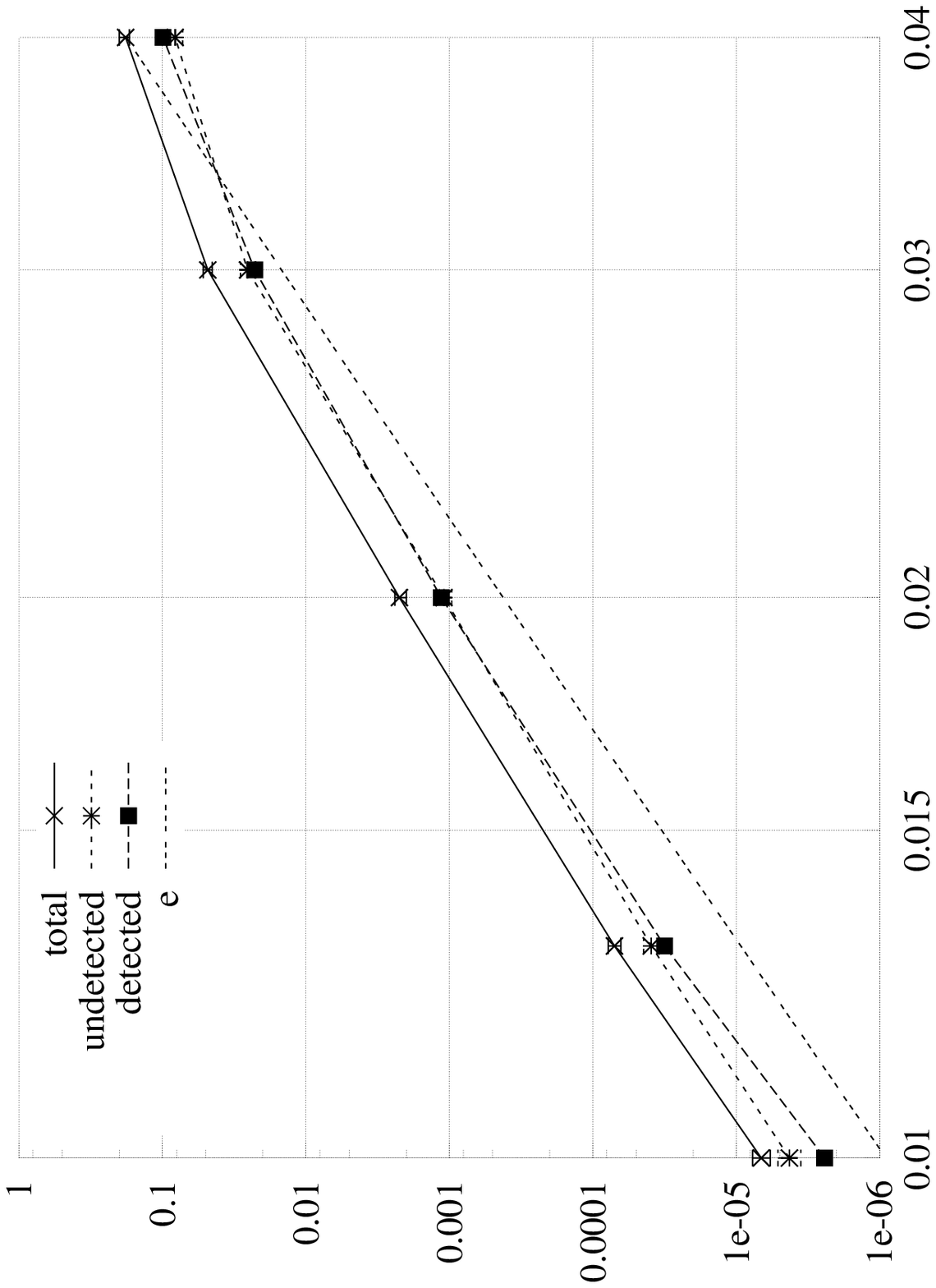}
\end{center}
\caption[a]{Performance of the construction-U code with $N=274$ and $M=82$
 on the binary symmetric channel,
 as a function of the flip probability $\fm$. The vertical axis shows
 the block error probability. Roughly half the errors were detected
 and half undetected. The three curves with error bars show the total block error rate,
 the detected error rate, and the undetected error rate. Also shown
 is an estimate $e$ of the error rate of the maximum likelihood
 decoder, assuming that the code has $A = 273^3$
 words of
 weight $d = 18$,  $e = A \, {{d}\choose{d/2}} f^{d/2} (1-f)^{d/2}$.
%
}
\label{fig.U}
\end{figure}
\subsection{Constructions U, N, and M}
 \Figref{fig.U} shows performance curves for the unicycle code with $N=274$.
 Whereas the codes of construction B presented thus far
 made no undetected errors, the codes of construction U
 sometimes gave undetected errors. To put it positively,
 the decoder for the code with $N=274$ performs almost
 as well as the maximum likelihood decoder for the code
 (all the errors made by a maximum likelihood decoder
 are undetected errors).
 The decoder also made detected errors, the frequency
 of errors of each type being roughly equal.
 Most of the undetected errors
 do not lie in the dual (so they are not admissible errors
 for the quantum code).

\begin{figure}
\begin{center}
\ebnobig{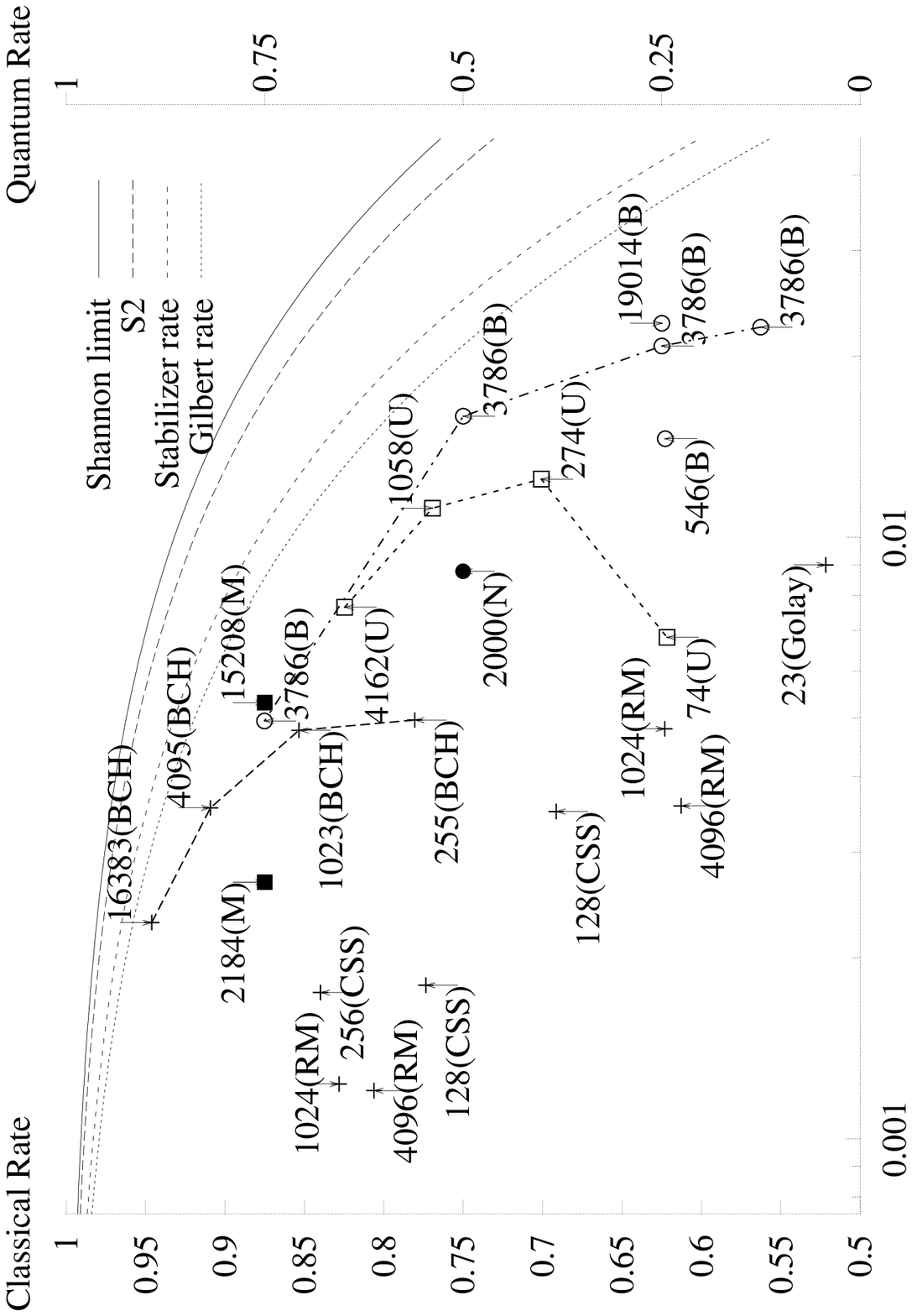}
\end{center}
\caption[a]{Summary of performances of several quantum codes
 on the 4-ary symmetric channel (depolarizing channel),
 treated (by all decoding algorithms shown in this figure)
 as if the channel were a pair of independent binary symmetric
 channels.
 Each point shows the  marginal
 noise level $\fm$ at which the block error probability
 is $10^{-4}$.
 In the case of dual-containing codes,
 this is the noise level at which each of the two identical
 constituent codes (see \eqref{eq.dualcon}) has an error probability of $5\times 10^{-5}$.

 As an aid to the eye, lines have been added between the four unicycle codes (U);
 between a sequence of bicycle codes (B) all of blocklength $N=3786$
 with different rates;
 and between a sequence of of BCH codes with increasing blocklength.

 The curve labelled S2 is the Shannon limit if the correlations
 between \X\ errors and \Z\ errors are neglected, (\ref{eq.CQBSC}).

 Points `$+$' are codes invented elsewhere. All other point styles
 denote codes presented for the first time in this paper.
}
\label{fig.QS1}
\end{figure}

 To make it easy to compare many quantum codes
 simultaneously, we summarise each code's performance by finding the
 noise level $\fm$ at which its block error probability
 is $10^{-4}$.
 In the case of a quantum code based on a dual-containing classical
 code, decoded by treating the channel
 as if it  were a pair of independent binary symmetric channels,
 the value plotted  is the noise level at which
 the block error probability of one constituent classical
 code is $0.5 \times 10^{-4}$.
 \Figref{fig.QS1} compares these performance summaries
 for a selection of codes.
 The right-hand vertical axis shows the quantum rate of the codes.
 The left-hand vertical axis shows the classical rate of the
 underlying classical code, where this concept is applicable.
 The figure  shows the performance of four Unicycle codes,
 two rate-$\dfrac{3}{4}$
 codes of construction M  and one rate-$\dhalf$ code of construction N.
 And it includes the performance of some algebraically-defined  codes,
 CSS codes based on BCH codes, Reed-Muller codes, and the Golay code.

\subsection{Exploiting channel knowledge when decoding}
\begin{figure}
\begin{center}
\ebnobig{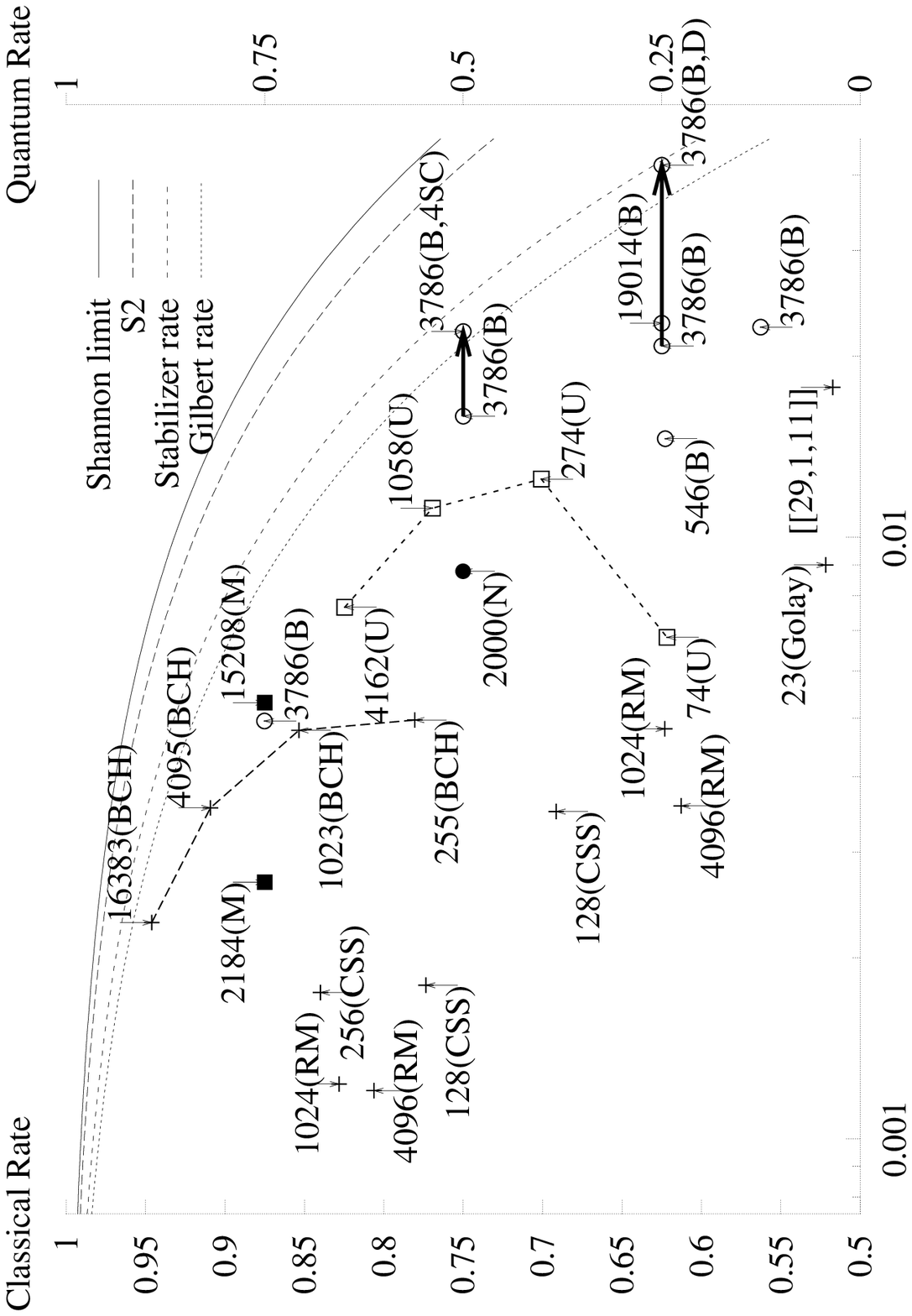}
\end{center}
\caption[a]{Summary of performances of several codes
 on the 4-ary symmetric channel (depolarizing channel).
 The additional points at the right and bottom are as follows.

 3786(B,4SC): a code of construction B (the same code as its
 neighbour in the figure) decoded with a decoder
 that exploits the known correlations between \X\ errors and \Z\ errors.

 3786(B,D):  the same code as the $N=3786$ code to its
 left in the figure, simulated with a channel where
 the qubits have a diversity
 of known reliabilities;   \X\ errors  and \Z\ errors occur independently
 with probabilities  determined from a Gaussian distribution; the channel
 in this case is not the 4-ary symmetric channel, but we
 plot the performance at the equivalent value of $\fm$.

 [[29,1,11]]: an algebraically constructed quantum code (not a sparse-graph code)
 from \protect\cite{grassl2003}.
}
\label{fig.QS2}
\end{figure}

 We can improve the performance of all the sparse-graph codes
 presented here by putting  knowledge about the channel properties
 into the  decoding algorithm.
 For example, if the channel is  the $4$--ary symmetric channel
 with error probability $q$, then instead of treating
 this channel as if it  consisted of two independent binary symmetric channels
 with flip probability $\fm = 2q/3$ (as is normal practice
 with CSS codes), we can build  knowledge of the correlations between \X\ errors and \Z\
 errors into the sum-product decoding algorithm.
\[
\begin{array}{c|cc}
\multicolumn{1}{c}{P(\noise_x,\noise_z)} & \multicolumn{1}{c}{\noise_x=0} & \noise_x=1 \\ \cline{2-3}
\noise_z=0      & 1\!-\!q & q/3 \\ 
\noise_z=1      & q/3     & q/3 
\end{array}
\]
 We illustrate the benefits of building this knowledge
 into the sparse graph's decoder by giving the results for
 one code in  \figref{fig.QS2}. There, we show the performance
 of a rate-$\dhalf$ bicycle code (classical rate 0.75) of length $N=3786$
 before (left point) and after inclusion of correlation knowledge (right point, marked `4SC').
 This code's performance is beyond the Gilbert rate and is therefore
 better than any performance that could be achieved by
 a traditional CSS code (that is, a CSS code composed of
 two binary codes each separately decoded by a bounded-distance decoder).


 The opportunity to build in channel knowledge gives 
 sparse-graph codes  a second major advantage  if the decoder knows
 that certain bits are more reliable than others.
 It is well known in communication theory that the ability to
 exploit such soft channel information can improve decoder
 performance by a couple of decibels, on standard benchmark channels.
 To illustrate this benefit, we took a classical rate-0.625
 bicycle code  of blocklength $N=3786$ and simulated its
 decoding  with the  channel likelihoods being generated
 using a Gaussian noise model as described in section \ref{sec.logit}.
 We show the performance of the decoder (by the point marked 3786(B,D)) in
 \figref{fig.QS2} at a horizontal coordinate $\fm$
 corresponding to the mean flip probability of the channel.

 \Figref{fig.QS2} also shows the performance of a
 small quantum code drawn from \cite{grassl2003}.
 As new quantum codes are invented, we intend to maintain
 a graph summarising the best performances on the depolarizing
 channel on the web at {\tt{www.inference.phy.cam.ac.uk/qecc/}}.
 Authors of quantum codes are encouraged to simulate their decoders
 and submit the results.

\section{Discussion}
 We have presented four families of dual-containing codes
 that are sparse-graph codes, all  based on cyclic matrices.

 The only parameter regimes in which
 we are aware of quantum codes of comparable blocklength that can surpass
 the sparse-graph codes presented here are the large-rate regime
 and the very small-rate regime:
 there exist sequences of dual-containing BCH codes
 with increasing blocklength $N$ and  rate $R_Q$ tending to 1
 \cite{Steane98}, which,
 as illustrated in \figref{fig.QS1},
 dominate at rates above $R_Q = 0.8$; and sequences of
 surface codes, with rates tending to zero, can cope with
 noise levels up to $\fm \simeq 0.11$ under the two-binary-symmetric-channels
 noise model \cite{DennisKitaevLandahlPreskill}. These surface codes
 are sparse-graph codes with very low-weight interactions.

 We have  estimated the performance of
 some of  Steane's `enlarged' codes \protect\cite{Steane98} and,
 although these have somewhat higher
 rate than the BCH codes they are derived from, they do not
 surpass the codes shown here.

\subsection{Encoding complexity}
 While all the sparse-graph codes presented have low {\em decoding\/}
 complexity, both in terms of the number of quantum interactions required
 and the number of classical operations required to infer the noise
 from the syndrome, 
 we have said nothing so far about their {\em encoding\/} complexities.
 The worst case is that the encoding complexity will scale 
 as $N^2$,
\ifnum \arabic{longversion} > 0 
 if we follow the recipe in \secref{sec.qencode}
 without exploiting the structure of the sparse graph.
\else
 if we follow standard encoding procedures
 without exploiting the structure of the sparse graph  (see \cite[Chapter 10]{NielsenChuang}
 and \cite{mackaymitchisonmcfadden2003Long}).
\fi
 However, in the case of constructions U, N, and M, all of which 
 directly involve sparse cyclic matrices, we think it is likely 
 that the cyclic structure can be exploited to yield encoders
 of lower complexity.
 For all four constructions B, U, N, and M, another option
 would be to try the methods of
\ifnum \arabic{longversion} > 0 
 \citeasnoun{Urbanke00}
\else
 Richardson and Urbanke \cite{Urbanke00}
\fi
 for lowering the complexity of encoding.
 We are not addressing the details of encoding at present as we
 are still hunting for even better codes; we will return
 to encoding
 when we have reached the limits of 
 our ability to generate promising sparse-graph codes.

\subsection{Concerning dual-containing codes}
 We hope that it is possible to find other constructions
 that might surpass these codes in terms of the parameters
 $R$ (rate), $\fm$ (noise level), or $k$ (sparseness of parity check matrix).

 Since finding pseudorandom dual-containing codes seems so
 difficult, it might be worthwhile
 to explore algebraically constructed sparse-graph codes.
 We  examined a dual-containing Euclidean Geometry code with blocklength $N=511$,
 kindly supplied by Shu Lin. This code's classical rate was 0.875 and
 it achieved a block error probability of $0.5 \times 10^{-4}$ at $\fm \simeq 0.00133$,
 a disappointing result; the code has low-weight codewords and
 at least  one tenth of the decoding errors were associated with these
 codewords.

 The decoding algorithm that we have used
 for all these codes is the plain sum-product algorithm.
 It is  known  this algorithm becomes increasingly suspect
 as the number of short cycles in the graph increases;
 and for dual-containing codes, the graph has an enormous
 number of cycles of length four.
 In the case of constructions B, N, and M, our decoder
 ignores these four cycles.
 It seems highly likely that
 a decoding algorithm that took these four-cycles
 into account would perform significantly better \cite{YFW2000,YFW2002}.
 Our  attempts to make such an
 improved algorithm have so far yielded only  algorithms whose
 complexity scales as $2^k$, where $k$ is the
 row-weight of the parity-check matrix.  Given that
 our preferred codes have $k \simeq 20$,
 these algorithms are regrettably not feasible.
 Construction U has an advantage here: while the code has four cycles
 as required, our decoder works by separately searching for decodings
 within two subcodes, both defined by graphs with no four-cycles.

\subsection{Prior work on sparse-graph codes for quantum error correction}
 Several constructions and decoding algorithms have been proposed for quantum codes
 that are associated with sparse graphs \cite{Kitaev97,Dennis00,DennisKitaevLandahlPreskill,Postol}.
 The major differences in our work are that we present codes with large blocklengths and with
 a wide variety of rates, and our codes can correct hundreds of errors;
 every sparse-graph code in the references listed above  either has vanishing rate ($K/N \simeq 0$ for large $N$)
 or cannot correct more than a tiny number of errors.

\subsection{Theoretical questions remaining}
 We think it would be very interesting to resolve
 a query about the distance properties of
 dual-containing codes with sparse parity check matrix.
 All the dual-containing codes we have found so far
 have the disappointing property that they have
 codewords (not in the dual) of weight $\leq k$,
 where $k$ is the row-weight of the parity check matrix.

 We offer two mutually exclusive conjectures about binary codes, one
  pessimistic   and one optimistic, identified
 by the initials of the author
 of each conjecture.
\begin{description}
\item[Conjecture G:]
 Any dual-containing
 code defined by an $M\! \times\! N$ parity check matrix $\bH$
 with $M<N/2$, all of whose rows have weight $\leq k$,
 has codewords of weight $\leq k$ that are not in the dual.
\item[Conjecture D:]
 There exist  dual-containing
 codes with sparse parity-check matrix and good distance.
 To be precise, such codes would have a parity-check matrix with maximum row weight
 $k$, and for increasing blocklength $N$
 the minimum distance $d$ of codewords not in the dual
 would satisfy $d \propto N$.
\end{description}

 If both these conjectures are false, we will be happy,
 because the middle-ground -- dual-containing sparse-graph codes
 with minimum distance $>k$ -- would be sufficient to give excellent
 practical performance.

\subsection{Beyond dual-containing codes}
 While we think the constructions reported here -- especially
 construction B -- are very promising and flexible, we hope 
 to find even better practical quantum codes.

 Having found the dual-containing constraint
 to be quite a severe one,
 we are now working on less-constrained sparse-graph codes,
 namely ones that satisfy the twisted product
 constraint (\ref{twistzero}) only.

\section*{Acknowledgments}
 We thank Shu Lin, Marc Fossorier, Mike Postol, and Andrew Landahl
 for helpful discussions. 
 DJCM is supported by the Gatsby Charitable Foundation
 and by a partnership award from IBM Z\"urich Research Laboratory.

\small\raggedright
\bibliography{bibs}

\end{document}